\documentclass[paper]{JHEP3}
\pdfoutput=1
\usepackage{amsmath,amssymb,amsthm,amscd,graphicx}
\input epsf.sty

\theoremstyle{definition}

\def\IP{{\mathbb P}}

\newcommand{\re}{{\rm e}}
\newcommand{\ri}{{\rm i}}
\newcommand{\rd}{{\rm d}}

\newcommand{\np}{{\mathrm{np}}}

\newcommand{\Li}{\mathop{\rm Li}\nolimits}

\newcommand{\be}{\begin{equation}}
\newcommand{\ee}{\end{equation}}
\newcommand{\ba}{\begin{aligned}}
\newcommand{\ea}{\end{aligned}}
\newcommand{\ben}{\begin{eqnarray}\displaystyle}
\newcommand{\een}{\end{eqnarray}}

\newcommand{\sectiono}[1]{\section{#1}\setcounter{equation}{0}}

\newcommand{\beq}{\begin{equation}} 
\newcommand{\eeq}[1]{\label{#1}\end{equation}} 
\newcommand{\bs}{\begin{split}} 
\newcommand{\es}{\end{split}} 

\newdimen\tableauside\tableauside=1.0ex
\newdimen\tableaurule\tableaurule=0.4pt
\newdimen\tableaustep
\def\phantomhrule#1{\hbox{\vbox to0pt{\hrule height\tableaurule width#1\vss}}}
\def\phantomvrule#1{\vbox{\hbox to0pt{\vrule width\tableaurule height#1\hss}}}
\def\sqr{\vbox{%
  \phantomhrule\tableaustep
  \hbox{\phantomvrule\tableaustep\kern\tableaustep\phantomvrule\tableaustep}%
  \hbox{\vbox{\phantomhrule\tableauside}\kern-\tableaurule}}}
\def\squares#1{\hbox{\count0=#1\noindent\loop\sqr
  \advance\count0 by-1 \ifnum\count0>0\repeat}}
\def\tableau#1{\vcenter{\offinterlineskip
  \tableaustep=\tableauside\advance\tableaustep by-\tableaurule
  \kern\normallineskip\hbox
    {\kern\normallineskip\vbox
      {\gettableau#1 0 }%
     \kern\normallineskip\kern\tableaurule}%
  \kern\normallineskip\kern\tableaurule}}
\def\gettableau#1{\ifnum#1=0\let\next=\null\else
\squares{#1}\let\next=\gettableau\fi\next}

\title{The spectral problem of the ABJ Fermi gas}

\author{
Johan K\"all\'en \\
D\'epartement de Physique Th\'eorique,\\
Universit\'e de Gen\`eve, Gen\`eve, CH-1211 Switzerland\\
\\
\email{johan.kallen@unige.ch}
}

\abstract{The partition function on the three-sphere of ABJ theory can be rewritten into a partition function of a non-interacting Fermi gas, with an accompanying one-particle Hamiltonian. We study the spectral problem defined by this Hamiltonian. We determine the exact WKB quantization condition, which involves quantities from refined topological string theory, and test it successfully against numerical calculations of the spectrum. 
}

\begin{document}

\sectiono{Introduction}
The partition functions of many superconformal Chern--Simons-matter theories can be reduced to matrix models through the process of localization \cite{kapustin}. Many of these theories have gravity duals given by M-theory on ${\rm AdS}_4\times X_7$, where $X_7$ is some seven dimensional manifold. A number of examples of this three-dimensional version of the AdS/CFT duality \cite{Maldacena:1997re} have been found, see for example \cite{Aharony:2008ug,Aharony:2008gk,Jafferis:2008qz}. If we want to use the matrix model describing the partition function in the gauge theory in order to learn about the gravity dual, we need to study them in a large $N$ expansion. Two different types of large $N$ expansions can be considered for these matrix models. The first one is the standard 't Hooft expansion, in which the coupling of the gauge theory scales with $N$. This type of large $N$ expansion probes the string theory regime of the large $N$ dual. To probe the M-theory regime, we need to instead study the matrix model in a large $N$ expansion in which the coupling of the gauge theory is held fixed. This type of expansion has been coined the \textit{M-theory expansion}, and matrix models which allows for this type of expansion are called \textit{M-theoretic matrix models.} In \cite{albanf}, general aspects of M-theoretic matrix models are discussed. By the nature of the AdS/CFT duality, the matrix models describing the partition functions of Chern--Simons-matter theories with gravity duals are M-theoretic matrix models, but there are also examples outside the context of gauge/gravity duality, given for example by \cite{Kazakov:1998ji,Kharchev:1992iv,Kostov:1995xw}. 

Standard methods to compute the full M-theory expansion of a given matrix model is today lacking. One way to approach the problem, which applies to a subset of the M-theoretic matrix models, was proposed in the seminal paper \cite{abjmfermi}. In there it was shown how to rewrite the matrix models for many superconformal  Chern--Simons-matter theories into the form of a partition function of a non-interacting quantum Fermi gas. In this description $N$ is interpreted as the number of particles in the gas, and the M-theory expansion corresponds to studying the thermodynamic limit of the Fermi gas. The Fermi gas picture defines a one-particle Hamiltonian.  In principle, all the information needed to compute the thermodynamic limit is encoded in the spectrum of this Hamiltonian. Therefore, computing the M-theory expansion of a matrix model which can be rewritten into a partition function of a Fermi gas can be reduced to finding the solution to a certain spectral problem. 

Of all the M-theoretic matrix models, the partition function of the so called ABJM model \cite{Aharony:2008ug} is the one which to date is best understood. The ABJM model is a superconformal Chern--Simons-matter theory with gauge group $U(N)\times U(N)$ and coupling constant $k$. It was noticed in \cite{marinoexact} that this matrix model is closely related to the matrix model description of the partition function of topological string theory on the Calabi--Yau manifold known as local $\IP^1 \times \IP^1$. Thanks to this relation it has been completely solved in the 't Hooft expansion \cite{weaktostrong}.  Furthermore, a series of works \cite{Hatsuda:2012hm,hmo2,hmo3,Calvo:2012du} led up to a conjecture of the full M-theory expansion of the partition function in \cite{nonpertstring}. In the M-theory expansion, both the standard free energy of the topological string on local $\IP^1\times \IP^1$ as well as the Nekrasov--Shatashvili limit of the refined topological string, on the same manifold, appears. The latter corresponds to effects which are invisible in the large $N$ 't Hooft expansion. Part of the works which laid the ground for the conjecture put forward in \cite{nonpertstring} was the exact computation of the partition function for various low-integer values of $(N,k)$ with subsequent extrapolation to large $N$ \cite{Putrov:2012zi,Hatsuda:2012hm,hmo2}. The spectral problem associated to the ABJM model was studied in detail in  \cite{qspec}. In there, it was solved through a WKB quantization condition and many of the conjectures in \cite{nonpertstring} regarding the partition function was understood in a new way, and some proven. Especially, the appearance of the Nekrasov--Shatashvili limit of the refined topological string is completely natural from this point of view.

A generalization of the ABJM model is the ABJ model \cite{Aharony:2008gk}. From the gauge theory perspective, the difference between the two models is simply that the two gauge groups are allowed to have different ranks in the ABJ model, whereas they must be the same for the ABJM model. When studying the 't Hooft expansion of the matrix model description of the partition functions, this small difference is of no technical importance. In fact, the most convenient way to study the ABJM matrix model in the 't Hooft expansion is to first allow for different ranks of the gauge groups and in the end set them equal. For the M-theory expansion the situation is different. To begin with, the method of rewriting the partition function of the ABJM matrix model into a Fermi gas found in \cite{abjmfermi} does not straightforwardly apply when the gauge groups have different ranks. However, in later works two different Fermi gas formalism for the partition function of the ABJ model has been found. In \cite{Matsumoto:2013nya} it is shown that the partition function of the ABJ model can be written as an expectation value of certain Wilson loops in the ABJM model. With a different approach, the papers \cite{Awata:2012jb,Honda:2013pea,Honda:2014npa} have shown that the partition function of the ABJ model can be factored into three different parts: the partition function of pure Chern--Simons theory on $S^3$, a phase and a partition function of a certain ideal quantum Fermi gas. Both the phase and the pure Chern--Simons theory partition function is known explicitly. Therefore the problem is once again reduced to study the thermodynamic limit of a Fermi gas, with a Hamiltonian which generalize the one of the ABJM Fermi gas. In \cite{Matsumoto:2013nya} and \cite{Honda:2014npa} the techniques of \cite{Putrov:2012zi,Hatsuda:2012hm,hmo2} are employed, which leads to conjectured forms of the partition function of the ABJ model along the lines of \cite{nonpertstring}.

The purpose of this paper is to study the spectral problem associated to the Fermi gas formalism found in \cite{Honda:2014npa}. We call this the \textit{spectral problem of ABJ theory}. As will be clear later on, even though the matrix models describing the ABJM and ABJ partition functions are very similar in character, the corresponding spectral problems are quite different. For example, for the ABJM model the connection to the refined topological string in the Nekrasov--Shatashvili limit is clear and an important aspect of the solution in \cite{qspec}. A priori, as will be further explained below, for the ABJ spectral problem there does not seem to be such a connection. Nevertheless, inspired by the results in  \cite{Matsumoto:2013nya,Honda:2014npa} we will argue that we can solve the ABJ spectral problem through a WKB quantization condition, where again topological strings play an important role. The expressions we obtain  generalize the ones in \cite{qspec}. We lack a proof of our claim, but we will test it with high precision by comparing the spectrum computed using the WKB method with numerical values of the spectrum. The agreement is excellent. We will also check that the partition function computed based on the results for the spectrum agrees with the expressions in  \cite{Matsumoto:2013nya,Honda:2014npa}.

The paper is organized as follows. In section \ref{abj} we will review the formulation of  the spectral problem. We will also review the solution to the spectral problem of the ABJM model in \cite{qspec} and conjecture how this solution is generalized for the ABJ spectral problem. In section \ref{wkbtest} we will perform tests of the conjecture against numerical values of the spectrum. In section \ref{abjpart} we will compute the M-theory expansion of the partition function of ABJ theory using our knowledge of the spectrum and compare with the corresponding expressions in \cite{Matsumoto:2013nya,Honda:2014npa}. We will end with conclusions and a discussion about how to approach other spectral problems appearing when computing the M-theory expansion of other Chern--Simons-matter theories. There are two appendices. A few details of a calculation in section \ref{qvolabjsec} can be found in appendix \ref{qvolapp}, whereas in appendix \ref{mellin} a review of the Mellin transform, used for the calculations in section \ref{abjpart}, can be found.

\section{The ABJ matrix model and the spectral problem} \label{abj}
The ABJ model \cite{Aharony:2008gk} is a $\mathcal{N}=6$ superconformal Chern--Simons-matter theory with gauge group $U(N_1)_k\times U(N_2)_{-k}$. The parameter $k$ is the Chern--Simons level, and it comes with opposite sign for the two gauge groups. The gravity dual of the theory is M-theory on the manifold ${\rm AdS}_4\times S^7/\mathbb{Z}_k$, with $|N_1-N_2|/k+1/2$ units of three-form flux through the three cycle $S^3/\mathbb{Z}_k$ \cite{Maldacena:1997re,Aharony:2008gk,Aharony:2009fc}. We will consider the gauge theory on the manifold $S^3$. The partition function of the theory can be reduced to a matrix model using localization \cite{kapustin}. We will call this matrix model the \textit{ABJ matrix model} and it is given by \cite{kapustin,weaktostrong}
\beq
\bs
&Z(N_1,N_2,k)=\frac{\ri^{-\frac{1}{2}\left(N_1^2-N_2^2\right)}}{N_1!N_2!} \\
&\times \int\prod_{i=1}^{N_1}\frac{\rd \mu_i}{2\pi}\prod_{j=1}^{N_2}\frac{\rd \nu_j}{2\pi}\frac{\prod_{i<j}\left(2\sinh{\left(\frac{\mu_i-\mu_j}{2}\right)}\right)^2\prod_{i<j}\left(2\sinh{\left(\frac{\nu_i-\nu_j}{2}\right)}\right)^2}{\prod_{i,j}\left(2\cosh{\left(\frac{\mu_i-\nu_j}{2}\right)}\right)^2} \re^{\frac{\ri k}{4\pi}\left(\sum_i \mu_i^2-\sum_j \nu_j^2\right)}
\end{split}
\eeq{zabj}  
For definiteness, we will in this paper without loss of generality assume that 
\beq
N_1\geq N_2
\eeq{}
and 
\beq
k\geq 0~.
\eeq{}
In \cite{Aharony:2008gk} it was argued that the ABJ model does not exists quantum mechanically as a unitary superconformal field theory unless $k$ fulfills the bound
\beq
k\geq N_1-N_2~,
\eeq{kn1n2}
which for $N_1\neq N_2$ sets a lower bound on $k$. 

The ABJ matrix model is an example of a matrix model which can be studied in two different expansions. First we have the standard 't Hooft expansion in which we study the model in the limit
\beq
N_1,~N_2,~k\rightarrow \infty
\eeq{}
while keeping the 't Hooft parameters $N_1/k$ and $N_2/k$ fixed. This limit of the model have been analyzed in detail in \cite{weaktostrong}. For the other type of expansion we introduce the parameters $N,M$ given by
\beq
N=N_1~,\quad \quad M=N_1-N_2~.
\eeq{NM} 
The \textit{M-theory expansion} of the ABJ matrix model is given by studying the limit
\beq
N\rightarrow \infty
\eeq{}
while keeping the parameters $k$ and $M$ fixed. As mentioned in the introduction, a general discussion of this type of expansion of a matrix model can be found in \cite{albanf}. 

In \cite{abjmfermi} a method to systematically study a certain type of matrix models in the M-theory expansion was introduced. Namely, if we can rewrite the matrix model into the form of a partition function of a one-dimensional quantum Fermi gas, it was shown in \cite{abjmfermi} how to use standard statistical mechanics techniques in order to compute the M-theory expansion. The Fermi gas formulation of the matrix model defines for us a one-particle Hamiltonian $\hat{H}$. The operator $\hat{H}$ defines a spectral problem, and all the information about the M-theory expansion of the matrix model is encoded in the solution of this spectral problem. In this paper, we are going to study the spectral problem associated to the matrix model $\eqref{zabj}$; the spectral problem of ABJ theory. This spectral problem has been introduced in \cite{Honda:2014npa}, building on the results in \cite{Awata:2012jb,Honda:2013pea}. Let us review how it is derived. 

Firstly, it is shown that the ABJ matrix model $\eqref{zabj}$ can be rewritten as
\beq
Z(N,N+M,k)=\re^{\ri \theta(N,M,k)}Z_{{\rm CS}}(M,k) \widehat{Z}(N,N+M,k)~.
\eeq{}
Above, $Z_{{\rm CS}}(M,k)$ is given by
\beq
Z_{{\rm CS}}(M,k)=k^{-M/2}\prod_{s=1}^{M-1}\left(2\sin{\left(\frac{\pi s}{k}\right)}\right)^{M-s}~.
\eeq{}
The subscript CS is due to that this is the partition function of pure Chern--Simons theory on $S^3$ with gauge group $U(M)$ (without the shift of the Chern--Simons level). For the values $M=0$ and $M=1$ the term involving the product is defined as 1, so we have 
\beq
Z_{{\rm CS}}(0,k)=1~,\quad \quad Z_{{\rm CS}}(1,k)=k^{-1/2}~.
\eeq{}
The phase $\theta(N,M,k)$ is given by
\beq
\re^{\ri \theta(N,M,k)}=\ri^{-(N^2+NM)}(-1)^{{N \over 2}(N-1)+NM}\ri^{N+NM}\re^{\frac{\ri \pi}{6}M(M^2-1)}~.
\eeq{}
Finally, $\widehat{Z}(N,N+M,k)$ can be written as
\beq
\widehat{Z}(N,N+M,k)=\frac{1}{N!}\sum_{\sigma\in S_N} \int \rd^{N}x \prod_{i=1}^N\rho(x_i,x_{\sigma(i)})
\eeq{zhatfermi}
where the function $\rho(x_1,x_2)$ is given by
\beq
\rho(x_1,x_2)=\frac{1}{2\pi k}\frac{\sqrt{U(x_1,M)U(x_2,M)}}{2\cosh{\left(\frac{x_1-x_2}{2k}\right)}}~,
\eeq{rhoabj}
with $U(x,M)$ given by
\beq
U(x,M)=\log{\left(\re^{x/2}+(-1)^M\re^{-x/2}\right)}-\sum_{m=-\frac{(M-1)}{2}}^{\frac{M-1}{2}}\log{\left(\tanh{\left(\frac{x}{2k}+\frac{\ri \pi m}{k}\right)}\right)}~.
\eeq{abjU}
The sum over $m$ in the above expression runs with the step $\Delta m=1$, and for $M=0$ the sum vanishes. 

The point of all this rewriting is that $\widehat{Z}(N,N+M,k)$ in $\eqref{zhatfermi}$ can be identified with the partition function of a one-dimensional Fermi gas with $N$ particles and one-particle density matrix in the position representation
\beq
\rho(x_1,x_2)=\langle x_1| \hat{\rho}| x_2\rangle
\eeq{}
given by $\eqref{rhoabj}$. The operator $\hat{\rho}$ defines the one-particle Hamiltonian $\hat{H}$ of the gas through 
\beq
\hat{\rho}=\re^{-\hat{H}}~.
\eeq{qhamabj}
In terms of the conjugate operators $\hat{x},\hat{p}$ fulfilling the canonical commutator relation
\beq
[\hat{x},\hat{p}]=\ri \hbar 
\eeq{}
we can write $\hat{\rho}$ as
\beq
\hat{\rho}=\re^{-U(\hat{x},M)/2}\re^{-T(\hat{p})}\re^{-U(\hat{x},M)/2}~,
\eeq{rhohatabj}
where $U(x,M)$ is given by $\eqref{abjU}$ and $T(p)$ is given by
\beq
T(p)=\log{\left(2 \cosh{\frac{p}{2}}\right)}~,
\eeq{abjT}
if we identify the Planck constant $\hbar$ with $k$ in the following way  
\beq
\hbar=2\pi k~.
\eeq{}
Therefore, the Chern--Simons level $k$ plays the role of the Planck constant in the Fermi gas treatment. As in \cite{qspec}, we will use $k$ and $\hbar$ interchangeably in this paper. We will call the function $H_{{\rm cl}}(x,p,M)$ given by 
\beq
H_{{\rm cl}}(x,p,M)=T(p)+U(x,M)
\eeq{hclabj}
the classical limit of $\hat{H}$ (even though the resulting expression still contains factors of $k$ for $M>0$).

The spectral problem of ABJ is defined by
\beq
\int_{-\infty}^{\infty} \rho(x_1,x_2)\phi(x_1)\rd x_1 = \re^{-E} \phi(x_2)
\eeq{inteqabj}
where $\phi(x)$ are normalizable functions and $\rho(x_1,x_2)$ is given by $\eqref{rhoabj}$. In order to have a well defined spectral problem we need, not surprisingly, to require that
\beq
k\geq M~,
\eeq{km}
just as is the case in the original formulation of ABJ theory. Only when $k$ and $M$ fulfill this condition, the potential energy of the gas $\eqref{abjU}$ is bounded from below, and the integral kernel $\eqref{rhoabj}$ defines a non-negative, Hermitian, Hilbert-Schmidt operator.

An alternative way to formulate the spectral problem is given by rewriting $\eqref{inteqabj}$ into the difference equation
\beq
\psi(x+\ri \pi k)+\psi(x-\ri \pi k)=\re^{-U(x,M)}\re^{E}\psi(x)~.
\eeq{diffeqabj}
Provided certain analyticity and boundary conditions for the function $\psi(x)$ is fulfilled the spectral problem $\eqref{inteqabj}$ is equivalent to the spectral problem defined by $\eqref{diffeqabj}$. Following \cite{Tracy:1995ax}, these conditions are given as follows. We denote by $\mathcal{S}_a$ the strip in the complex $x$-plane defined by 
\beq
|{\rm Im}(x)|<a~.
\eeq{}
Let us also denote by $A(\mathcal{S}_a)$ those functions $g$ which are bounded and analytic in the strip, continuous on its closure, and for which $g(x+\ri y)\rightarrow 0$ as $x\rightarrow \pm \infty$ through real values, when $y\in \mathbb{R}$ is fixed and satisfies $|y|<a$. Using the results of  \cite{Tracy:1995ax} it can be seen that $\eqref{inteqabj}$ and $\eqref{diffeqabj}$ are equivalent if $\psi(x)$ belongs to the space $A(\mathcal{S}_{\pi k})$. 

We notice that the spectral problems for $M=0$ and $M>0$ are quite different in character. For $M>0$, there are explicit factors of $k$ in the potential energy $\eqref{abjU}$. Such potentials have also appeared in other Fermi gas matrix models, see for example \cite{{abjmfermi},{albanf},{Mezei:2013gqa}}. In addition, we have a lower bound on $k$, given by equation $\eqref{km}$. Since $k$ plays the role of $\hbar$ in our quantum Fermi gas, from a physical point of view these features are unusual. 

 For the special case of $M=0$, the spectral problem was studied in detail in \cite{qspec}. For the original matrix model, the case of $M=0$ corresponds to what is known as the ABJM model \cite{Aharony:2008ug}. In the approach of \cite{qspec}, the spectrum is found using the so called WKB method.  In this approach, the energy levels $E_n$ are determined by the \textit{WKB quantization condition}
\beq
{\rm vol}(E;\hbar,M)=2\pi \hbar\left(n+\frac{1}{2}\right)~,\quad n=0,1,2,\ldots~,
\eeq{wkbcond}
where ${\rm vol}(E;\hbar,M)$ is the \textit{quantum volume of phase space}. Below, we will review how  ${\rm vol}(E;\hbar,0)$ is found in \cite{qspec}. Due to the difference in the character of the spectral problems it is not obvious that there is a function ${\rm vol}(E;\hbar,M)$ which solves the spectral problem $\eqref{inteqabj}$ through the condition $\eqref{wkbcond}$ also for $M>0$. However, we will in this paper give a proposal for such a function.
\subsection{Review of the solution of the spectral problem for $M=0$}
In general, the quantum volume has two different parts. One has a perturbative expansion in $\hbar$, and we denote it by ${\rm vol}_{{\rm p}}(E;\hbar,0)$. The other one is non-perturbative in $\hbar$, meaning that it involves terms which are non-analytic at  $\hbar=0$, and we denote it by ${\rm vol}_{{\rm np}}(E;\hbar,0)$. So we have
\beq
{\rm vol}(E;\hbar,0)={\rm vol}_{{\rm p}}(E;\hbar,0)+{\rm vol}_{{\rm np}}(E;\hbar,0)~.
\eeq{}
As will be further explained below, expanding the LHS of $\eqref{wkbcond}$ to lowest order in a small $\hbar$ expansion the WKB quantization condition reduces to the well known Bohr--Sommerfeld quantization condition. The perturbative $\hbar$ corrections was first written down in \cite{dunham}. Papers addressing the problem of computing non-perturbative corrections to the WKB quantization condition includes \cite{ZinnJustin:1981dx,ZinnJustin:2004ib,ZinnJustin:2004cg,MR729194,MR1483488,MR1704654}. 

The building blocks in computing both  ${\rm vol}_{{\rm p}}(E;\hbar,0)$ and ${\rm vol}_{{\rm np}}(E;\hbar,0)$ are period integrals on the curve in phase space defined by the equation
\beq
\re^{H_{{\rm cl}}(x,p,0)}=\re^{E}~.
\eeq{abjmcurve} 
As was noticed in \cite{abjmfermi}, this curve is a specialization of the curve describing the mirror of the Calabi--Yau manifold known as local $\IP^{1}\times\IP^{1}$. The equation for this curve is usually written as
\beq
\re^{u}+ z_1 \re^{-u}+\re^{v}+z_2 \re^{-v}=1~,
\eeq{}
where $u,v$ are complex coordinates and $z_{1},z_{2}$ are the two complex structure parameters of the mirror Calabi--Yau. If we make a change of variables
\beq
u=\frac{x+p}{2}-E~,\quad \quad v=\frac{x-p}{2}-E
\eeq{}
we see that the two curves are the same if we identify of the complex structure parameters and the energy $E$ in the following way
\beq
z_1=z_2=z~,
\eeq{z1z2Ecl}
where we for convenience have introduced the notation
\beq
z=\re^{-2E}~.
\eeq{}
In the quantum theory the identification of $z_{1},z_{2}$ with the Fermi gas parameters involves a quantum correction \cite{qspec}
\beq
z_1=q^{1/2}z~,\quad \quad z_2=q^{-1/2}z~,
\eeq{z1z1Eq}
where
\beq
q=\re^{\ri \pi k}~.
\eeq{}
This provides a link between the spectral problem $\eqref{diffeqabj}$ for $M=0$ and topological string theory on local $\IP^1\times\IP^1$.

The perturbative part of the quantum volume can be computed as follows. Make an ansatz for the solution of $\eqref{diffeqabj}$ of the form
\beq
\psi(x)=\re^{\frac{1}{\hbar} S(x,\hbar)}
\eeq{}
where $S(x,\hbar)$ has a $\hbar$ expansion of the form
\beq
S(x,\hbar)=\sum_{n\geq 0} S_n(x) \hbar^{n}~.
\eeq{}
The function ${\rm vol}_{{\rm p}}(E;\hbar,0)$ is then given by
\beq
{\rm vol}_{{\rm p}}(E;\hbar,0)=\oint_{\gamma} \partial_x S(x,\hbar)~,
\eeq{volpint}
where $\gamma$ is a cycle on the curve $\eqref{abjmcurve}$ around the two turning points defined by the solutions to the equation
\beq
H_{{\rm cl}}(x,0,0)=E~.
\eeq{}
The classical limit
\beq
\hbar\rightarrow 0
\eeq{}
of $\eqref{volpint}$ is given by
\beq
\lim_{\hbar \to 0}{\rm vol}_{{\rm p}}(E;\hbar,0)=\oint_{\gamma} p(x,E)\rd x~,
\eeq{vol0}
where $p(x,E)$ is obtained by solving $\eqref{abjmcurve}$. The integral on the RHS in $\eqref{vol0}$ calculates the volume enclosed by the contour $\gamma$ in phase space. This is why we call the function ${\rm vol}(E;\hbar,0)$ the quantum volume of phase space; it is given by the classically available volume of phase space for a given energy together with quantum corrections. 

The period integral $\eqref{vol0}$ is closely related to what is usually called the B period in the topological string theory literature. To obtain the quantum corrections, that is, to compute the quantum B period, one can follow the standard prescription, which consists of calculating order by order in a small $\hbar$ expansion. However, in the context of the ABJM model it would be much desired to instead calculate the quantum B period for \textit{fixed} $\hbar$, but in an expansion for large $E$. This corresponds, in the original matrix model, to the M-theory expansion, that is, to a large $N$ expansion for fixed $k$. In \cite{qgeom}, a method to compute quantum periods for fixed $\hbar$, but as an expansion in the complex structure parameters, was introduced. This method can be applied to the mirror curve of local $\IP^1\times \IP^1$. Furthermore, in \cite{qgeom,Mironov:2009uv,Mironov:2009dv} it was shown that the perturbative quantum B period is closely related to the free energy of the refined topological string in the Nekrasov--Shatashvili limit \cite{Nekrasov:2009rc}. See also \cite{Huang:2014nwa} for a discussion about computations of quantum periods.

For general complex structure parameters, there are two different quantum B periods, which we denote by $\Pi_{B_{I}}(z_1,z_2;\hbar)$, $I=1,2$. They are related by an exchange of moduli 
\beq
\Pi_{B_{1}}(z_1,z_2;\hbar)=\Pi_{B_{2}}(z_2,z_1;\hbar)
\eeq{}
and can be written
\beq
\bs
\Pi_{B_1}(z_1, z_2;\hbar)&=-{1\over 8}\left( \log^2 z_1  -2 \log z_1\log z_2 -\log^2 z_2 \right) +{1\over 2} \log z_2\, \widetilde \Pi_A (z_1, z_2;\hbar)\\
&+ {1\over 4} \widetilde \Pi_B (z_1, z_2;\hbar)~,
\end{split}
\eeq{}
where $\widetilde{\Pi}_{A}(z_1,z_2;\hbar)$ and $ \widetilde \Pi_B (z_1, z_2;\hbar)$ can be computed systematically in a power series in $z_{1},z_2$ \cite{{qgeom,nonpertstring}}. As shown in \cite{qspec}, the combination of quantum $B$ periods which gives the perturbative part of the quantum volume is
\beq
\bs
{\rm vol}_{{\rm p} }(E; \hbar,0)&= 4 \Pi_{B_1} ( q^{1/2} z, q^{-1/2} z; \hbar) + 4 \Pi_{B_2} ( q^{1/2} z, q^{-1/2} z; \hbar)  -{4 \pi^2 \over 3} -{\hbar^2 \over 12}\\
&= 8 E^2 -{4 \pi^2 \over 3} +{\hbar^2 \over 24} - 8 E \sum_{\ell\ge 1}  \widehat a_\ell (\hbar) \re^{-2 \ell E} + 2 \sum_{\ell\ge1} \widehat b_{\ell } (\hbar) \re^{-2 \ell E}
\end{split}
\eeq{volpabjm}
where the coefficients $\widehat{a}_\ell(\hbar)$ and $\widehat{b}_\ell(\hbar)$ are defined by
\beq
\bs
&\widetilde \Pi_A (q^{1/2}z, q^{-1/2}z;\hbar)=\sum_{\ell\geq 1} \widehat{a}_\ell(\hbar) z^\ell~, \\
&\frac{1}{2}\left(\widetilde \Pi_B (q^{1/2}z, q^{-1/2}z;\hbar)+\widetilde \Pi_B (q^{-1/2}z, q^{1/2}z;\hbar)\right)=\sum_{\ell\geq 1} \widehat{b}_\ell(\hbar) z^\ell~.
\end{split}
\eeq{pitilde}
For the non-perturbative part of the quantum volume, ${\rm vol}_{{\rm np}}(E;\hbar,0)$, by general principles it is instead the quantum A period $\Pi_{A_I}(z_1,z_2;\hbar)$ that appears \cite{MR729194,MR1483488,MR1704654,abjmfermi,qspec}.  As for the B periods, there are two A periods. They are given by
\beq
\Pi_{A_I}(z_1,z_2;\hbar)=\log{z_I}+\widetilde{\Pi}_A(z_1,z_2;\hbar)~,\quad \quad I=1,2~.
\eeq{} 
To calculate the non-perturbative part of the quantum volume for the ABJM spectral problem from first principles is a difficult, unsolved problem. However, in \cite{qspec} it is conjectured that ${\rm vol}_{{\rm np}}(E;\hbar,0)$ is closely related to the standard, un-refined, topological string free energy on local $\IP^1 \times \IP^1$. In Gopakumar-Vafa form, this quantity is given by \cite{gv}
\beq
F(T_1,T_2,g_s)=\sum_{g\geq 0}\sum_{m\geq 1}\sum_{m|d}\sum_{d_1+d_2=d}\frac{d}{m}n_g^{d_1,d_2}\left(2\sin{\left(\frac{\ri g_s m}{2d}\right)}\right)^{2g-2}\re^{-\frac{m}{d}\left(d_1 T_1+d_2 T_2\right)}~.
\eeq{}
In the above formula, $g_s$ is the topological string coupling constant, $T_{1,2}$ are the complexified K\"ahler classes and $n_g^{d_1,d_2}$ are the Gopakumar-Vafa invariants of local $\IP^1\times \IP^1$. In \cite{qspec} it is conjectured that ${\rm vol}_{{\rm np}}(E;\hbar,0)$ is given by
\beq
\bs
{\rm vol}_{{\rm np}}(E;\hbar,0)=-4\pi k \sum_{g\geq 0}\sum_{m\geq 1}\sum_{m|d}\sum_{d_1+d_2=d}&\sin{\left(\frac{4\pi k}{m}\right)}\frac{d}{m} n_g^{d_1,d_2}\left(2\sin{\left(\frac{2\pi m}{dk}\right)}\right)^{2g-2} \\
&\times \re^{-\frac{m}{dk}\left(d_1 \Pi_{A_1}(q^{1/2}z,q^{-1/2}z;\hbar)+d_2\Pi_{A_2}(q^{1/2}z,q^{-1/2}z;\hbar)\right)}~.
\end{split}
\eeq{volnpabjm}
A major inspiration for the conjecture $\eqref{volnpabjm}$ comes from the fact that there are coefficients in the large energy expansion of the perturbative part of the quantum volume that diverges for rational values of $k$. However, the original spectral problem $\eqref{inteqabj}$ is perfectly well defined for these values of $k$.  With the non-perturbative part of the quantum volume given by $\eqref{volnpabjm}$ all poles cancel out, so the total function ${\rm vol}(E;\hbar,0)$ does not have poles for any real value of $\hbar$. This is an implementation of the so called Hatsuda--Moriyama--Okuyama cancelation mechanism discovered in  \cite{hmo2}. 

In summary, using the fact that the constant energy surface in phase space $\eqref{abjmcurve}$ can be identified with the mirror curve of local $\IP^1\times \IP^1$, the perturbative part of the quantum volume for $M=0$ can be calculated from first principles and is given $\eqref{volpabjm}$, whereas the non-perturbative part has the conjectured form $\eqref{volnpabjm}$.
\subsection{The quantum volume for $M>0$} \label{qvolabjsec}
The goal of this paper is to find the function ${\rm vol}(E;\hbar,M)$ also for $M> 0$. At first sight the WKB method seems somewhat problematic for $M>0$. This is since we have a lower bound on $\hbar$, and a WKB quantization procedure is usually associated with the existence of a small $\hbar$ expansion. A problem from a practical point of view is that, for $M>0$, it does not seem like we can identify the constant energy surface
\beq
\re^{H_{{\rm cl}}(x,p,M)}=\re^{E}
\eeq{curveabj}
with the mirror curve of local $\IP^1\times \IP^1$ for any choice of identification of the complex structure parameters and the energy, as we could for $M=0$. For these reasons, it seems hopeless to approach the spectral problem $\eqref{inteqabj}$ with the WKB method, and even if we could, it does not seem that we can get any help from results in topological string theory. On the other hand, as can be seen from the analysis in the 't Hooft limit \cite{weaktostrong}, the ABJ matrix model is clearly related to topological string theory on local $\IP^1 \times \IP^1$, also for $N_1\neq N_2$. We would therefore expect that there is a connection to topological strings for the spectral problem appearing in the Fermi gas formulation, also for $M>0$. 

A very useful method to get a first hand on the problem of calculating the quantum volume in a large $E$ expansion given a classical Hamiltonian was found in \cite{abjmfermi}.  This method is based on the Wigner approach to quantization \cite{Hillery:1983ms}, and it neglects terms involving $\re^{-E}$ and $\re^{-E/\hbar}$, but otherwise gives exact results in $\hbar$. It has been applied in for example \cite{abjmfermi,Mezei:2013gqa} when the "classical" Hamiltonian depends on $\hbar$, just as in our case for $M>0$. Applying the method with the Hamiltonian given by  $\eqref{qhamabj}$ and $\eqref{rhohatabj}$ we find
\beq
\bs
{\rm vol}(E;\hbar,M)=&8 E^2 -{4 \pi^2 \over 3} +{\hbar^2 \over 24} +2\pi^2\left(M-\frac{\hbar}{2\pi}\right)M +\mathcal{O}(\re^{-E},\re^{-E/\hbar})~.
\end{split}
\eeq{volpabj0}
The calculation of the above result can be found in appendix \ref{qvolapp}. We do not know how to systematically calculate the exponentially small corrections, but we notice that taking the combination of quantum B periods given in the first line of $\eqref{volpabjm}$, but with the complex structure parameters instead given by  

\beq
\bs
z_1&=\rm{e}^{-2E+\frac{\rm{i}\pi k}{2}-\rm{i}\pi M} \\
z_2&=\rm{e}^{-2E-\frac{\rm{i}\pi k}{2}+\rm{i}\pi M}~,
\end{split}
\eeq{z1z2}
we reproduce the displayed terms in $\eqref{volpabj0}$. Encouraged by this result, we conjecture that the quantum volume ${\rm vol}(E;\hbar,M)$ which solves the ABJ spectral problem using the WKB quantization condition is given by the sum of $\eqref{volpabjm}$ and $\eqref{volnpabjm}$, but where the arguments of the quantum periods are given by the RHS of $\eqref{z1z2}$ instead:
 \beq
{\rm vol}(E;\hbar,M)={\rm vol}_{{\rm p}}(E;\hbar,M)+{\rm vol}_{{\rm np}}(E;\hbar,M)
\eeq{qvolabj}
where
\beq
\bs
{\rm vol}_{{\rm p}}(E;\hbar,M)=& 4 \Pi_{B_1} \left(\re^{-\ri \pi M} q^{1/2} z, \re^{\ri \pi M}q^{-1/2} z; \hbar\right) + 4 \Pi_{B_2} \left( \re^{-\ri \pi M} q^{1/2} z, \re^{\ri \pi M}q^{-1/2} z; \hbar\right)  \\
& -{4 \pi^2 \over 3} -{\hbar^2 \over 12}\\
=& 8 E^2 -{4 \pi^2 \over 3} +{\hbar^2 \over 24} +2\pi^2\left(M-\frac{\hbar}{2\pi}\right)M \\
&- 8 E \sum_{\ell\ge 1} (-1)^{M\ell} \widehat  a_\ell (\hbar) \re^{-2 \ell E} + 2 \sum_{\ell\ge1}  (-1)^{M\ell}\widehat b_{\ell } (\hbar) \re^{-2 \ell E}
\end{split}
\eeq{volpabj}
and
\beq
\bs
{\rm vol}_{{\rm np}}(E;\hbar,M)&=-4\pi k \sum_{g\geq 0}\sum_{m\geq 1}\sum_{m|d}\sum_{d_1+d_2=d}\sin{\left(\frac{4\pi k}{m}\right)}\frac{d}{m} n_g^{d_1,d_2}\left(2\sin{\left(\frac{2\pi m}{dk}\right)}\right)^{2g-2} \\
&\times \re^{-\frac{m}{dk}\left[d_1 \Pi_{A_1}\left(\re^{-\ri \pi M} q^{1/2}z,\re^{\ri \pi M} q^{-1/2}z;\hbar\right)+d_2\Pi_{A_2}\left(\re^{-\ri \pi M} q^{1/2}z,\re^{\ri \pi M}  q^{-1/2}z;\hbar\right)\right]} \\
&=-4\pi k\sum_{m\geq 1} \sin{\left(\frac{4\pi m}{k}\right)}d_m(k,M)\re^{-4mE_{{\rm eff}}/k}~.
\end{split}
\eeq{volnpabj}
In the last line we have introduced the notation\footnote{Note that $d_m(k,0)$ differs from $d_m(k)$ in \cite{qspec} by a factor $(-1)^m$.}
\beq
\bs
d_m(k,M)&=\sum_{g\ge 0} \sum_{d|m}\sum_{d_1+d_2=d} {d   \over m}\,(-\beta^{-1})^{d_1m/d}(-\beta)^{d_2m/d} \,n_g^{d_1,d_2}  \left( 2 \sin {2 \pi m \over d k} \right)^{2g-2} ~,\\
\beta&=\re^{-2\pi \ri M/k}
\end{split}
\eeq{betadm}
and 
\beq
E_{{\rm eff}}=E-\frac{1}{2}\sum_{\ell\geq 1}(-1)^{M\ell} \widehat{a}_{\ell}(\hbar)\re^{-2\ell E}~.
\eeq{eeff}
In the next section we will collect strong evidence that this is the correct quantum volume for the ABJ spectral problem by calculating energy levels using the WKB quantization condition and compare against numerical values obtained directly from $\eqref{inteqabj}$. Before we do this, let us mention two other aspects of the above expressions which supports the claim that it is the correct quantum volume; invariance under a Seiberg like duality and cancelation of poles.
\subsubsection{Seiberg like duality}
In \cite{Aharony:2008gk} it is argued that the ABJ theory with gauge groups 
\beq
U(N+M)_k\times U(N)_{-k}
\eeq{}
and 
\beq
U(N)_k\times U(N+k-M)_{-k}
\eeq{}
give equivalent theories. In \cite{Kapustin:2010mh,Willett:2011gp}, this Seiberg like duality was checked using the ABJ matrix model. In the Fermi gas approach, it translates to the invariance of the Hamiltonian defining the ABJ spectral problem under
\beq
M\to k-M~,
\eeq{seiberg}
as shown  in \cite{Honda:2014npa}. A necessary condition for the quantum volume is therefore that it is invariant under $\eqref{seiberg}$. That the function $\eqref{qvolabj}$ fulfills this requirement can be seen as follows. Under the transformation $\eqref{seiberg}$ the complex structure parameters given in $\eqref{z1z2}$ are exchanged 
\beq
z_1\leftrightarrow z_2~.
\eeq{}
The $M$-dependence in $\eqref{qvolabj}$ comes solely from the complex structure parameters $\eqref{z1z2}$ and since the transformation $\eqref{seiberg}$ exchange these parameters, the perturbative part of the quantum volume is invariant by construction. For the non-perturbative part, the $M$-dependence is in $\beta$ and $E_{{\rm eff}}$.  The $\beta$ in $\eqref{betadm}$ is clearly invariant under the Seiberg like duality and that $E_{{\rm eff}}$ is invariant can be seen using $\eqref{pitilde}$ together with the fact that $\widetilde{\Pi}_{A}(z_1,z_2)$ is a symmetric function in $z_1$ and $z_2$. 
\subsubsection{Pole cancelation}
Also for $M>0$, both the perturbative and the non-perturbative part of the quantum volume have poles. In this section we will check that the poles cancel out. As in \cite{qspec}, we rewrite both ${\rm vol}_{{\rm p}}(E;\hbar,M)$ and  ${\rm vol}_{{\rm np}}(E;\hbar,M)$ in terms of the refined BPS invariants $N_{j_L,j_R}^{d_1,d_2}$ of local $\IP^1\times \IP^1$ \cite{Iqbal:2007ii}. First, we re-express the perturbative volume in terms of the variable $E_{{\rm eff}}$, given in equation $\eqref{eeff}$. The function  ${\rm vol}_{{\rm p}}(E;\hbar,M)$ can then be written as 
\beq
{\rm vol}_{{\rm p}}(E;\hbar,M)=8E_{{\rm eff}}^2  -{4 \pi^2 \over 3} +{\hbar^2 \over 24} +2\pi^2\left(M-\frac{\hbar}{2\pi}\right)M +4\pi^2 k\sum_{\ell\geq 1}(-1)^{M\ell}\widetilde{b}_\ell(k)\re^{-2\ell E_{{\rm eff}}}~,
\eeq{volpbps}
where $\widetilde{b}_\ell(k)$ are the same coefficients as in \cite{qspec,nonpertstring}. In terms of $N_{j_L,j_R}^{d_1,d_2}$ they are given by \cite{nonpertstring}
\be
\widetilde{b}_\ell(k)=-\frac{\ell}{2\pi}\sum_{j_L,j_R}\sum_{\ell=dw}\sum_{d_1+d_2=d}N^{d_1,d_2}_{j_L,j_R}q^{\frac{w}{2}(d_1-d_2)}
\frac{\sin\frac{\pi kw}{2}(2j_L+1)\sin\frac{\pi kw}{2}(2j_R+1)}{w^2\sin^3\frac{\pi kw}{2}}~. 
\ee
Next we use that the Gopakumar-Vafa invariants $n_{g}^{d_1,d_2}$ can be related to the refined BPS invariants $N_{j_L,j_R}^{d_1,d_2}$. The relevant relations can be found in \cite{nonpertstring}. Using this we find that  $\eqref{betadm}$ can be written
\beq
\bs
d_m(k,M)&=\sum_{j_L,j_R} \sum_{m=dn}\sum_{d_1+d_2=d} {1   \over n}\,(-\beta^{-1})^{d_1m/d}(-\beta)^{d_2m/d} \,N_{j_L,j_R}^{d_1,d_2} \frac{2j_R+1}{\left(\sin {2 \pi n \over  k} \right)^2}\frac{\sin {\left({4 \pi n \over  k}(2j_L+1)\right)} }{\sin {4 \pi n \over  k} }~.
\end{split}
\eeq{}
There are poles in both ${\rm vol}_{{\rm p}}(E;\hbar,M)$ and ${\rm vol}_{{\rm np}}(E;\hbar,M)$ for any rational value of $k$. Let us verify to the poles cancel between the two expressions. For a given rational value of $k$, let us consider the terms in $\eqref{volpbps}$ and $\eqref{volnpabj}$ with summation indices fulfilling 
\beq
k=\frac{2n}{w}=\frac{2m}{\ell}~.
\eeq{krational}
Expanding the term in ${\rm vol}_{{\rm p}}(E;\hbar,M)$ around this value of $k$ we find
\beq
(-1)^{M\ell+1}\re^{\ri \pi kw (d_1-d_2)/2}\frac{8m}{w^3\left(k-\frac{2n}{w}\right)} (-1)^{n(2j_L +2 j_R-1)}(1+2j_L) (1+2j_R) N_{j_L,j_R}^{d_1,d_2}\re^{-2\ell E_{{\rm eff}}}
\eeq{}
whereas the term in ${\rm vol}_{{\rm np}}(E;\hbar,M)$ has the pole structure
\beq
\re^{\frac{2\pi \ri M}{k}m(d_1-d_2)/d}(-1)^m \frac{8m}{w^3\left(k-\frac{2n}{w}\right)}(1+2j_L) (1+2j_R) N_{j_L,j_R}^{d_1,d_2}\re^{-\frac{2 m w}{n} E_{{\rm eff}}}~,
\eeq{}
which generalizes the pole structure of the quantum volume derived for $M=0$ in \cite{qspec} to $M\geq 0$. A geometric argument explained in \cite{qspec} gives that 
\beq
(-1)^{n(2j_L +2 j_R-1)}=1
\eeq{}
and using this together with $\eqref{krational}$ we find that the poles present for rational values of $k$ cancel between ${\rm vol}_{{\rm p}}(E;\hbar,M)$ and  ${\rm vol}_{{\rm np}}(E;\hbar,M)$ for all $M\geq 0$. 
\section{Testing the WKB quantization condition} \label{wkbtest}
In the previous section we proposed that the eigenvalue problem $\eqref{inteqabj}$ is solved by the WKB quantization condition $\eqref{wkbcond}$ using the quantum volume $\eqref{qvolabj}$.  Although we are not able to prove this fact, we will in this section perform a detailed test of it. The most obvious way to test a proposed solution of a quantum problem is in a small $\hbar$ expansion. For $M>0$, due to the condition $\eqref{km}$, this is not possible in this case. In \cite{qspec}, a way to test the WKB quantization condition for \textit{finite} $\hbar$ was used. Since the quantum volume is given by a large $E$ expansion for fixed $\hbar$, we can calculate the energy levels in a large quantum number expansion, for fixed $\hbar$. For the case $M=0$ it was shown in \cite{qspec} that this method gave good values already for the lowest lying energy levels. The eigenvalues computed in this way could then be compared with values obtained from solving $\eqref{inteqabj}$ numerically. For $M>0$, we can use the same approach to test if $\eqref{qvolabj}$ gives the correct quantum volume.

In general, good numerical values for the eigenvalues of $\eqref{inteqabj}$ is easier to obtain for low integer values of $(k,M)$. We will focus on testing the WKB quantization condition for $k=2,3$.  When $k=2,3$, the large $E$ expansion of ${\rm vol}(E;\hbar,M)$ is an expansion in $\re^{-4E/k}$. Let us write the quantum volume, for $k=2,3$, as
\beq
{\rm vol}(E;\hbar,M)=8E^2+\alpha(k,M)-8E\sum_{\ell\geq 1} A_\ell(\hbar,M) \re^{-4\ell E/k}+2\sum_{\ell\geq 1} B_\ell(\hbar,M) \re^{-4\ell E/k}
\eeq{qvolkfin}   
where
\beq
\alpha(k,M) = -{4 \pi^2 \over 3} +{\pi^2 k^2 \over 6} +2\pi^2(M-k)M ~.
\eeq{}
We notice that, for a given $k$, for those terms in the sum such that $2\ell/k$ is an integer, $A_\ell(\hbar)$ and $B_\ell(\hbar)$ will in general get contributions from both $\eqref{volpabj}$ and $\eqref{volnpabj}$. 
We assume an ansatz for the solution for the energy levels $E_n$ of the form
\beq
E_n=E_n^{(0)}+\sum_{\ell\geq 1}E_n^{(\ell)} \re^{-4\ell E_n^{(0)}/k}~.
\eeq{enexp}
Plugging this into $\eqref{qvolkfin}$ and using the WKB condition $\eqref{wkbcond}$ we can solve for $E_n^{(l)}$ recursively in $\ell$. To lowest order we find
\beq
\bs
E_n^{(0)}=&\sqrt{\frac{\pi\hbar}{4}\left(n+{1 \over 2}\right)-\frac{\alpha(k,M)}{8}}~,\\
\end{split}
\eeq{elev0}
which is valid provided $n$ is large enough. For $\ell \geq 1$ we have\footnote{We suppress the arguments of $A_\ell(\hbar,M)$ and $B_\ell(\hbar,M)$  for notational convenience.}
\beq
\bs
E_n^{(\ell)}=&\frac{1}{2E_n^{(0)}}\Bigg[E_n^{(0)} A_\ell -\frac{1}{4}B_\ell +\sum_{m=1}^{\ell-1}E_n^{(m)}A_{\ell-m}-\sum_{m=1}^{\ell-1} E_{n}^{(m)}E_{n}^{(\ell-m)} \\
&+\sum_{s=1}^{\ell-1}\left(\sum_{s\leq r+\sum_{q=2}^s m_q\leq \ell-1 }\left(\frac{-4r}{k}\right)^s\frac{1}{s!}E_n^{(\ell-\sum_{q=2}^s m_q-r)}E_n^{(m_2)}\cdots E_n^{(m_s)}\left( E_n^{(0)} A_{r}-\frac{1}{4}B_{r}\right) \right) \\
&+\sum_{s=1}^{\ell-2}\left(\sum_{s+1\leq r+\sum_{q=2}^s m_q+t\leq \ell-1 }\left(\frac{-4r}{k}\right)^s\frac{1}{s!}E_n^{(\ell-\sum_{q=2}^s m_q-r-t)}E_n^{(m_2)}\cdots E_n^{(m_s)}E_n^{(t)}A_r\right) \Bigg]~.
\end{split}
\eeq{elev1}
Next we evaluate the quantum volume for $k=2,3$. For the perturbative part of the quantum volume we need the coefficients $\widehat{a}_\ell(\hbar),\widehat{b}_\ell(\hbar)$; these can be found in for example \cite{nonpertstring,qspec}. For the non-perturbative part we need the Gopakumar-Vafa invariants $n_g^{d_1,d_2}$  of local $\IP^1\times \IP^1$; these are listed up to genus $g=8$ and total degree $d_1+d_2=10$ in for example \cite{Aganagic:2002qg}. Using these results we find, including the first few non-trivial exponentially small terms,
\beq
\bs
{\rm vol}(E;4\pi,M)=&8E^2-\frac{2\pi^2}{3}+2(M-2)M\pi^2+ 64 (-1)^ME\re^{-2E}+16(-1)^M \re^{-2E} +\ldots\\
{\rm vol}(E;6\pi,M)=&8E^2+\frac{\pi^2}{6}+2(M-3)M\pi^2 +8\sqrt{3}\pi \cos{\frac{2M\pi}{3}}\re^{-4E/3} \\
&+4\sqrt{3}\pi\left(2+\cos{\frac{4M\pi}{3}}\right)\re^{-8E/3}  + 32 E \re^{-4E}+4\re^{-4E} \\&-\frac{2\pi\left(21+67\cos{\frac{4M\pi}{3}}\right)}{\sqrt{3}}\re^{-16E/3}+\ldots\\
\end{split}
\eeq{}
From these expressions we can read off $A_\ell(\hbar,M),B_\ell(\hbar,M)$ and then use $\eqref{elev0}$ and $\eqref{elev1}$ to calculate the energy levels. The results for $M=1$ and the two lowest lying energy levels are displayed in table \ref{k2M1table} and \ref{k3M1table} .
\begin{table}[t] 
\centering
\begin{tabular}{c c c}
&Energy levels for $k=2,M=1$&\\
\hline
Order &	$E_0$ & $E_1$ \\
\hline
$0$ & $\underline{2.8}6786860477$ & \underline{4.25}373656158\\
$1$ & $\underline{2.881}90835982$ & \underline{4.254591}85227 \\
$2$ & $\underline{2.88181}489768$ & \underline{4.254591528}48\\
$3$ & $\underline{2.8818154}3241$ & \underline{4.25459152858}\\
\hline
Numerical value & 2.88181542992 &4.25459152858
\end{tabular}
\caption{The lowest and next-to-lowest energy eigenvalues for $k=2$, $M=1$ calculated analytically, including higher and higher orders of exponentially small corrections in $\eqref{enexp}$. In the last line numerical values are given obtained from the integral equation $\eqref{inteqabj}$ are given. At each order of the approximation, we underline the digits which agree with the numerical result. }
 \label{k2M1table}
\end{table}%
\begin{table}[t] 
\centering
\begin{tabular}{c c c}
&Energy levels for $k=3,M=1$&\\
\hline
Order &	$E_0$ & $E_1$ \\
\hline
$0$ & $\underline{3.48}301431852$ & \underline{5.1}8997064969\\
$1$ & $\underline{3.4867}7116444$ & \underline{5.190229}58113 \\
$2$ & $\underline{3.48669}611439$ & \underline{5.19022910}159\\
$3$ & $\underline{3.4866953}1759$ & \underline{5.1902291000}9\\
\hline
Numerical value & 3.48669532933 &5.19022910008
\end{tabular}
\caption{The lowest and next-to-lowest energy eigenvalues for $k=3$, $M=1$ calculated analytically, including higher and higher orders of exponentially small corrections in $\eqref{enexp}$. In the last line numerical values obtained from the integral equation $\eqref{inteqabj}$ are given. At each order of the approximation, we underline the digits which agree with the numerical result. }
 \label{k3M1table}
\end{table}%
We will now compare these analytically calculated energy levels with the numerical values obtained from the integral equation that defines our spectral problem, equation $\eqref{inteqabj}$. For $M=0$, it was shown in \cite{Hatsuda:2012hm} how to rewrite the integral equation $\eqref{inteqabj}$ into an eigenvalue equation for an infinite dimensional matrix ${\bf M}^{(k,M)}$. Since the kinetic energy $T(p)$ for our Fermi gas (given in equation $\eqref{abjT}$) does not depend on $M$,  we can use the same approach also for $M>0$. Following the derivation in \cite{Hatsuda:2012hm} we find that the matrix ${\bf M}^{(k,M)}$ is given by
\beq
{\bf M}_{ij}^{(k,M)}=\frac{1}{4\pi}\int_{-1}^1 t^{i+j} \exp{\left[U\Big(2k ~{\rm arctanh}(t),M\Big)\right]} \rd t
\eeq{}
where $U(x,M)$ is the potential energy of the Fermi gas, given in equation $\eqref{abjU}$. Explicitly, for $k=2$, $M=1$ and $k=3$, $M=1$ we find
\beq
\bs
{\bf M}_{ij}^{(2,1)}& =\left\{
  \begin{array}{l l}
   \frac{1}{2\pi(i+j+1)(i+j+3)}  & \quad \quad\quad\quad\quad \quad\quad\quad\quad\quad  \text{if $i+j$ is even}\\
    0 & \quad \quad\quad\quad\quad \quad\quad\quad\quad\quad \text{if $i+j$ is odd}
  \end{array} \right.  \\
{\bf M}_{ij}^{(3,1)}&=\left\{
  \begin{array}{l l}
  \frac {\Gamma \left(\frac{1}{2} (i+j+1)\right) \, _2F_1\left(1,\frac{1}{2} (i+j+1);\frac{1}{2}
   (i+j+6);-\frac{1}{3}\right)}{16
   \sqrt{\pi }}  & \quad \text{if $i+j$ is even}\\
    0 & \quad \text{if $i+j$ is odd}
  \end{array} \right.  ~.
\end{split}
\eeq{}
In order to compute the eigenvalues of ${\bf M}^{(k,M)}_{ij}$ with high numerical accuracy we follow the same approach as in \cite{qspec}. First off, as noticed in \cite{Hatsuda:2012hm}, since the matrix ${\bf M}^{(k,M)}$ 
has the following form
\be
{\bf M}^{(k,M)}=\begin{pmatrix}
m_0 & 0 & m_1 & 0 & m_2 & 0 & \ldots \\
0 & m_1 & 0 &  m_2 &0 & m_3  & \\
m_1 & 0 &  m_2 &0 & m_3  & 0 &  \\
0 &  m_2 &0 & m_3  & 0 & m_4 & \\
m_2 & 0 & m_3  & 0 & m_4 & 0 & \\
\vdots &  & & &  &  &\ddots
\end{pmatrix}~,
\ee
we can instead compute the eigenvalues of the matrices ${\bf M}^{(k,M)}_+$ and ${\bf M}^{(k,M)}_-$ given by:
\begin{align}
{\bf M}^{(k,M)}_+= \begin{pmatrix}
m_0 & m_1 & m_2 & \ldots \\
m_1 &  m_2 & m_3  & \\
m_2 & m_3 & m_4  \\
\vdots &    &  &\ddots
\end{pmatrix}, \qquad  & {\bf M}^{(k,M)}_-= \begin{pmatrix}
m_1 & m_2 & m_3 & \ldots \\
m_2 &  m_3 & m_4  & \\
m_3 & m_4 & m_5  \\
\vdots &    &  &\ddots
\end{pmatrix}~.
\end{align}
The eigenspaces of ${\bf M}^{(k,M)}$ decompose into a direct product of the eigenspaces of ${\bf M}^{(k,M)}_\pm$. Let the eigenvalues of ${\bf M}^{(k,M)}$ be denoted by $\lambda_n$, ordered such that
\be
\lambda_0 >\lambda_1 >\lambda_2 >\ldots~,
\ee 
and let the eigenvalues of ${\bf M}^{(k,M)}_\pm$ be denoted by $\lambda_{\pm,n}$, ordered in the same way. We then have
\begin{align}
\lambda_{+,n}=\lambda_{2n}, \quad \lambda_{-,n}=\lambda_{2n+1}~.
\end{align}
The relation between the eigenvalues of ${\bf M}^{(k,M)}$ and the energy eigenvalues is
\be
E_n=-\log{\lambda}_n~.
\ee

 Second off, in practice we have to truncate the infinite dimensional matrix to a finite $L\times L$ matrix. The eigenvalues of the truncated matrix $E_n(L)$ give numerical approximations of the exact eigenvalues $E_n$, and they converge to $E_n$ as $L\to\infty$. To accelerate the convergence we use Richardson extrapolation; see for example \cite{MR1721985} for an explanation of this method.
 
Using this approach we have calculated the energy eigenvalues numerically for the first few integer values of $k$ and $M$. The results for $k=2$, $M=1$ and $k=3$, $M=1$ are displayed in table \ref{k2M1table} and   \ref{k3M1table}, respectively. As we can see, the agreement between the numerical values obtained from $\eqref{inteqabj}$ and the analytical values calculated using the WKB quantization condition and the quantum volume $\eqref{qvolabj}$ is excellent. This strongly supports that we have found the correct quantum volume.
 \section{The partition function of ABJ} \label{abjpart}
The original motivation for studying the spectral problem $\eqref{inteqabj}$ is to compute the M-theory expansion of the ABJ matrix model. In the Fermi gas formulation of the matrix model the central object to compute is the grand potential $J(\mu,k,M)$, defined in terms of canonical partition function $\widehat{Z}(N,N+M,k)$ as
\beq
J(\mu,k,M)=\log{\left(1+\sum_{N=1}\widehat{Z}(N,N+M,k)\re^{N \mu}\right)}~.
\eeq{} 
If we know the spectrum of the one-particle Hamiltonian $\hat{H}$ of the Fermi gas, we can compute $J(\mu,M,k)$ by the formula
\beq
J(\mu,k,M)=\sum_{n\geq 0} \log{\left(1+\re^{\mu-E_n}\right)}~,
\eeq{jsum}
where $E_n$, $n=0,1,2,\ldots$ are the eigenvalues of $\hat{H}$. As shown in \cite{qspec}, using the Euler-Maclaurin formula we can rewrite the sum $\eqref{jsum}$ into an integral
\beq
J(\mu,k,M)=\frac{1}{2\pi \hbar }\int_{E_0}^\infty\frac{{\rm vol}(E;\hbar,M) \rd E}{\re^{E-\mu}+1}-\mathcal{R}(k,M)~.
\eeq{voltoj}
In the expression above, $E_0$ is the ground state energy of the Fermi gas and ${\rm vol}(E;\hbar,M)$ is the quantum volume determined in section \ref{qvolabjsec}. The function $\mathcal{R}(k,M)$ is defined by
\beq
\mathcal{R}(k,M)=\sum_{r\geq 1}\frac{B_{2r}}{(2r)!}f^{(2r-1)}(0)
\eeq{R} 
where $B_r$ is the $r$-th Bernoulli number and the function $f(n)$ is given by
\beq
f(n)=\log{\left(1+\re^{\mu-E(n)}\right)}~,
\eeq{}
where we have used the WKB condition $\eqref{wkbcond}$ to define a function $E(n)$ for arbitrary values of $n$. $f^{(r)}(n_0)$ denotes the $r$-th derivative of $f(n)$, evaluated at $n=n_0$. To compute the integrals appearing in $\eqref{voltoj}$ we can use the same method as was used for the case $M=0$ in \cite{qspec}. In there, a variant of the Mellin transform considered in \cite{ZinnJustin:2004cg} is used. In appendix \ref{mellin} we compute the relevant integrals. The final expression we obtain for $J(\mu,k,M)$ is quite complicated. Namely, using the results in appendix \ref{mellin} we find after a little rewriting that for $M\geq 0$ the grand potential $J(\mu,k,M)$ is given by
\beq
\bs
J(\mu,k,M)=&\frac{2}{3 \pi^2 k}\mu^3+B(k,M)\mu \\
& +\sum_{\ell\geq 1}\left(\frac{(-1)^{M\ell+1}\widehat{a}_\ell(\hbar)}{\pi^2 k}\mu^2 +\frac{(-1)^{M\ell}\widehat{b}_\ell(\hbar)}{\pi^2 k}\mu\right)\re^{-2\ell\mu} +\sum_{m\geq 1} d_m(k,M)\re^{-4 m \mu_{{\rm eff}}/k}\\
&+\widetilde{A}(k,M) +\sum_{\ell\geq 1} \widetilde{c}_\ell(\hbar,M) \re^{-2\ell \mu} +\sum_{\ell\geq 0}\widetilde{d}_\ell(\hbar,M)\re^{-(2\ell+1)\mu} ~.
\end{split}
\eeq{jfull}
In the above expression, $B(k,M)$ is given by
\beq
B(k,M)=\frac{1}{3k}+\frac{k}{24}+\frac{M}{2k}(M-k)~,
\eeq{}
and $\mu_{{\rm eff}}$ is given by
\beq
\mu_{{\rm eff}}=\mu-\frac{1}{2}\sum_{\ell\geq 1} (-1)^{M\ell} \widehat{a}_\ell(\hbar)\re^{-2\ell \mu}~.
\eeq{} 
In order to write the functions with a tilde in a somewhat compact way we introduce the following notation. By writing the non-perturbative part of the quantum volume as
\beq
{\rm vol}_{{\rm np}}(E;\hbar,M)=\sum_{m\geq 1}\sum_{r\geq 0}s_{r,m}(k,M)\re^{-\left(\frac{4m}{k}+2r \right)E}~,
\eeq{}
we define the coefficients $s_{r,m}(k,M)$. In addition we introduce
\beq
\bs
\mathcal{R}_0(k,M)&=-\sum_{r\geq 1}\frac{B_{2r}}{(2r)!}\frac{\rd^{2r-1}}{\rd n^{2r-1}}E(n)|_{n=0}~,\\
\mathcal{R}_\ell(k,M)&=\frac{(-1)^{\ell+1}}{\ell}\sum_{r\geq 1}\frac{B_{2r}}{(2r)!}\frac{\rd^{2r-1}}{\rd n^{2r-1}}\re^{\ell E(n)}|_{n=0}~,~\ell\geq 1~,
\end{split}
\eeq{Rl} 
as well as defining
$\mathcal{I}_j(n)$  by
\beq
\mathcal{I}_j(n)=\int_{E_0}^\infty E^j \re^{-n E}\rd E=\left(-\frac{\partial}{\partial n}\right)^j\left(\frac{1}{n} \re^{-nE_0}\right)~.
\eeq{cali}
$\mathcal{R}_\ell(k,M)$ is the coefficient of the term $\re^{-\ell\mu}$ in the expansion of $\eqref{R}$ at large $\mu$. With this notation the functions with a tilde are given by
\beq
\bs
\widetilde{A}(k,M)&=\widehat{A}(\hbar,M)-\mathcal{R}_0(k,M)  +\frac{1}{4\pi^2} \sum_{m\geq 1}\sum_{r\geq 0} \frac{s_{r,m}(k,M)}{4m+2kr} \re^{-\left(\frac{4m}{k}+2r \right)E_0} \\
\widetilde{c}_\ell(\hbar,M)&=\frac{\widehat{c}_\ell(\hbar,M)}{2\pi^2 k}-\mathcal{R}_{2\ell}(k,M)  +\frac{\re^{2\ell E_0}}{4\pi^2} \sum_{m\geq 1}\sum_{r\geq 0} \frac{s_{r,m}(k,M)}{4m+k(2r-2\ell)} \re^{-\left(\frac{4m}{k}+2r \right)E_0} \\
\widetilde{d}_\ell(\hbar,M)&= \frac{\widehat{d}_\ell(\hbar,M)}{2\pi^2 k}-\mathcal{R}_{2\ell+1}(k,M)  -\frac{\re^{(2\ell+1) E_0}}{4\pi^2} \sum_{m\geq 1}\sum_{r\geq 0} \frac{s_{r,m}(k,M)}{4m+k(2r-2\ell-1)} \re^{-\left(\frac{4m}{k}+2r \right)E_0}
\end{split}
\eeq{}
where 
\beq
\bs
\widehat{A}(\hbar,M)&=-\frac{1}{\pi\hbar}\left(\frac{4}{3} E_0^3+\frac{1}{2}\alpha(k,M)E_0+4\sum_{\ell \geq 1} (-1)^{M\ell}\widehat{a}_\ell(\hbar)\mathcal{I}_1(2\ell)-\sum_{\ell \geq 1} (-1)^{M\ell}\widehat{b}_\ell(\hbar)\mathcal{I}_0(2\ell)\right)\\
\widehat{c}_\ell(\hbar,M)&=-\frac{2\pi^2}{3} (-1)^{M\ell}\widehat{a}_\ell(\hbar) +2 (-1)^{M\ell}\widehat{a}_\ell(\hbar)E_0^2 -(-1)^{M\ell} \widehat{b}_\ell(\hbar)E_0 \\
&-4 \sum_{m\neq \ell}(-1)^{M m}\widehat{a}_m(\hbar)\mathcal{I}_1(2(m-\ell)) +\sum_{m\neq \ell}(-1)^{M m}\widehat{b}_m(\hbar)\mathcal{I}_0(2(m-l)) \\
&+4 \mathcal{I}_2(-2l)+\frac{1}{2}  \alpha(k,M)\mathcal{I}_0(-2l)~,\\
\widehat{d}_\ell(\hbar,M)&=4 \sum_{m\neq \ell}(-1)^{M m}\widehat{a}_m(\hbar)\mathcal{I}_1(2m-2\ell-1) -\sum_{m\neq \ell}(-1)^{M m}\widehat{b}_m(\hbar)\mathcal{I}_0(2m-2l-1) \\
&-4\mathcal{I}_2(-2l-1)-\frac{1}{2} \alpha(k,M)\mathcal{I}_0(-2l-1)~.
\end{split}
\eeq{}
The expression $\eqref{jfull}$ is obviously quite formidable. We now want to compare this expression with the expressions for the grand potential of ABJ theory obtained previously in the papers \cite{Matsumoto:2013nya,Honda:2014npa}. We notice that the first two lines on the RHS in $\eqref{jfull}$ does not depend on the ground state energy $E_0$, whereas the terms involving the functions with a tilde do. For the terms which does not depend on $E_0$, we can easily  compare with the corresponding terms in \cite{Matsumoto:2013nya,Honda:2014npa}, and we see that they match. Since we do not have a closed form expression for the ground state energy, it is harder to compare the other terms. In \cite{qspec}, for $M=0$ the first few of the other terms were checked in perturbation theory around $k=0$. Due to the lower bound on $k$, given in equation $\eqref{km}$, we cannot do the same type checks for $M>0$. However, from the previous section we know how to calculate the energy levels for  $k=2,3$, in a large quantum number expansion. The terms with a tilde can then be calculated in a similar expansion. More precisely, since we can compute the energy levels in the expansion $\eqref{enexp}$, the terms that depend on $E_0$ can be calculated in an expansion in $\re^{-4E_0^{(0)}/k}$. As we will see, similarly as for the energy levels the expansion seems to converge quite fast and we are able to obtain approximate values with high accuracy\footnote{The author would like to thank Yasuyuki Hatsuda for discussions about how these kind of checks for finite $k$ of the coefficients can be performed.}. We can then compare with the coefficients of the expressions in \cite{Matsumoto:2013nya,Honda:2014npa}. Before we perform these test we will discuss a type of term which was not present when calculating the energy levels, namely $\mathcal{R}_{\ell}(k,M)$, defined in $\eqref{Rl}$. 
\subsection{Borel resummation of $\mathcal{R}_\ell(k,M)$}
$\mathcal{R}_\ell(k,M)$ is in general a divergent series, since the Bernoulli numbers asymptotically grows as
\beq
B_{2r}\sim (2r)!~,\quad \quad r\to\infty
\eeq{}
and the factor with the $r$-th derivative in $\eqref{Rl}$ is not divided by a $r!$. For example, for  $\ell=0$, $k=2$, $M=1$ and to lowest order in the large quantum number expansion we have 
\beq
\mathcal{R}_0(2,1)=\pi\sum_{r\geq 1}\frac{B_{2r}}{(2r)!}\binom{1/2}{2r-1}(2r-1)!\left( {1 \over 2} +{\pi^2 \over 3}\right)^{-(2r-1)/2}~,
\eeq{}
and the sum on the RHS diverges. We can use Borel resummation in order to evaluate these divergent series. We will only give a brief description of this technique, for a pedagogical review see for example \cite{Marino:2012zq,Caliceti:2007ra}. Let us denote 
\beq
\bs
a_{r,0}&=-\frac{B_{r}}{r!}\frac{\rd^{r-1}}{\rd n^{r-1}}E(n)|_{n=0}~,\\
a_{r,\ell}&=\frac{(-1)^{\ell+1}}{\ell}\frac{B_{r}}{r!}\frac{\rd^{r-1}}{\rd n^{r-1}}\re^{\ell E(n)}|_{n=0}~,~\ell\geq 1~.
\end{split}
\eeq{}
The series  
\beq
\varphi_\ell(w)=\sum_{r\geq 1} a_{2r,\ell} w^{2r}
\eeq{}
has zero radius of convergence. Its \textit{Borel transform} $B_\ell(p)$ is defined by
\beq
B_\ell(p)=\sum_{r\geq 1}a_{2r,\ell}  \frac{p^{2r-1}}{(2r-1)!}~.
\eeq{} 
$B_\ell(p)$ typically defines a function which is analytic in a neighborhood of the origin. If we can analytically continue $B_\ell(p)$ to a neighborhood of the real line in such a way that the Laplace transform
\beq
\int_{0}^\infty \re^{-w p}B_\ell(p)\rd p
\eeq{}
is convergent the function $s(\varphi_\ell)(w)$ defined by
\beq
s(\varphi_\ell)(w)=\int_{0}^\infty \re^{-w p}B_\ell(p)\rd p
\eeq{}
is called the \textit{Borel sum} of $\varphi_\ell(w)$. Evaluating $s(\varphi_\ell)(w)$ at $w=1$ gives the Borel sum of $\eqref{Rl}$. 

For our expressions, even though we know explicitly all the coefficients in the Borel transform $B_\ell(p)$, we will not be able to evaluate it in closed form. However, we can apply the technique of \textit{Pad\'e approximants} in order to get an accurate approximation to the analytical continuation of $B_\ell(p)$. Given a power series
\beq
g(z)=\sum_{n\geq 0} a_n z^n 
\eeq{}
the Pad\'e approximant $[l/m]_g(z)$ is given by a ratio of two polynomials of degree $l$ and $m$, respectively:
\beq
[l/m]_g(z)=\frac{p_0+ p_1 z + p_2 z^2 +\ldots +p_l z^l}{1+ q_1 z + q_2 z^2 +\ldots +q_m z^m}~.
\eeq{pade} 
The original series $g(z)$ and the Pad\'e approximant $[l/m]_g(z)$ agrees up to order $l+m$:
\beq
g(z)-[l/m]_g(z)=\mathcal{O}(z^{m+l+1})~.
\eeq{} 
This equation determines the coefficients in $\eqref{pade}$. In the calculations in section \ref{coefffink} we will use a Pad\'e approximant to the series $B_\ell(p)$ of the form
\beq
\mathcal{P}^{(q)}_\ell(p)=\Big[[q/2]/[(q+1)/2]\Big]_{B_\ell}(p)~,
\eeq{npade}
where the notation $[x]$ means the integral part of  $x$, for various $q$. We will call the value of $q$ the degree of the Pad\'e approximant. Let us define $\mathcal{R}^{(q)}_{\ell}(k,M)$ by
\beq
\mathcal{R}^{(q)}_\ell(k,M)=\int_{0}^\infty \re^{- p}\mathcal{P}^{(q)}_\ell(p)\rd p~.
\eeq{}
$\mathcal{R}^{(q)}_\ell(k,M)$ gives an approximation to the Borel resummation of $\eqref{Rl}$. The approximation can be systematically improved by increasing $q$. 

See also \cite{resurEM} where expressions for the Borel transform of $\eqref{R}$ are obtained which might be useful in this context.
\subsection{Comparing coefficients for integer values of $k$} \label{coefffink}
As mentioned above, in \cite{Matsumoto:2013nya,Honda:2014npa} a proposal for the grand potential of ABJ theory is given. The expression in these papers is given by 
\beq
J(\mu,k,M)=\frac{2}{3 \pi^2 k}\mu^3+B(k,M)\mu+A(k,M) +J^{{\rm np}}(\mu,k,M)
\eeq{}
where we have used the definition of $J^{{\rm np}}(\mu,k,M)$ as given in \cite{Matsumoto:2013nya}. The $\mu$-independent term $A(k,M)$ is given by\footnote{The grand potential is defined slightly differently in \cite{Honda:2014npa} and \cite{Matsumoto:2013nya}, with the effect that the $\mu$-independent term differs by a term $|\log{Z_{CS}|}$ between the two papers. In this paper we use the definition for the grand potential given in \cite{Honda:2014npa}.}
\beq
\bs
A(k,M)=&-\frac{\zeta(3)}{8\pi^2}k^2 +\frac{1}{2}\log{2}+\frac{1}{6}\log{\frac{\pi}{2k}}+2 \zeta'(-2)-\frac{1}{3}\int_0^\infty \rd x \frac{1}{\re^{k x}-1}\left(\frac{3}{x^3}- \frac{1}{x}-\frac{3}{x \sinh^{2}{x}}\right)\\
&-\log{|Z_{CS}(k,M)|}~
\end{split}
\eeq{}
and in \cite{Matsumoto:2013nya} explicit expressions for the large $\mu$ expansion of $J^{{\rm np}}(\mu,k,M)$ for various integer $k$ is written down. We quote the first few terms for $k=2,3$, $M=1$ below:
\beq
\bs
J^{{\rm np}}(\mu,2,1)&=\left[-\frac{4\mu^2+2\mu}{\pi^2}-\frac{1}{\pi^2}\right]\re^{-2\mu} +\ldots\\
J^{{\rm np}}(\mu,3,1)&=-\frac{2}{3}\re^{-4\mu/3}-\re^{-8\mu/3}+\ldots \\ 
\end{split}
\eeq{jmorifink}

In order to compare our expressions for integer $k$ with the above formulas we first use the form of the quantum volume in $\eqref{qvolkfin}$ in the integration formula which determines $J(\mu,k,M)$, equation $\eqref{voltoj}$. Using the Mellin transform we find that, for $k=2,3$,  $J(\mu,k,M)$ is given by 
\beq
J(\mu,k,M)=\frac{2}{3 \pi^2 k}\mu^3+B(k,M)\mu +\widetilde{A}(k,M) +\widetilde{J}^{{\rm np}}(\mu,k,M) 
\eeq{}
where
\beq
\bs
\widetilde{J}^{{\rm np}}(\mu,k,M)=&\sum_{\ell\geq 1,\atop {4\ell \over k}\in \mathbb{N}}(-1)^{4\ell/k}\left(-\frac{A_\ell(\hbar,M)}{\pi^2 k}\mu^2 +\frac{B_\ell(\hbar,M)}{2\pi^2 k}\mu \right) \re^{-4\ell \mu/k} +\sum_{\ell\geq 1}\widetilde{C}_\ell(\hbar,M) \re^{-\ell \mu} \\
&+\sum_{\ell\geq 1,\atop {4\ell \over k}\notin \mathbb{N}}\left[\frac{2A_\ell(\hbar,M)}{\pi k}\left(\csc{\left(\frac{4\pi \ell}{k}\right)}\mu +\pi \cot{\frac{4\pi \ell}{k}}\csc{\frac{4\pi \ell}{k}}\right)-\frac{B_\ell(\hbar,M)}{2\pi k}\csc{\frac{4\pi \ell}{k}}\right]\re^{-4\ell \mu/k}\\
\end{split}
\eeq{jfink}
where now
\beq
\bs
\widetilde{A}(k,M)&=\widehat{A}(k,M) -\mathcal{R}_0(k,M)~, \\
\widetilde{C}_\ell(\hbar,M) & =\widehat{C}_\ell(\hbar,M) -\mathcal{R}_\ell(k,M)\end{split}
\eeq{}
with
\beq
\bs
\widehat{A}(k,M)&= -\frac{1}{\pi\hbar}\left(\frac{4}{3} E_0^3+\frac{1}{2}\alpha(k,M)E_0+4\sum_{\ell \geq 1} A_\ell(\hbar,M)\mathcal{I}_1(4\ell/k)-\sum_{\ell \geq 1} B_\ell(\hbar,M)\mathcal{I}_0(4\ell/k)\right) \\
\widehat{C}_\ell(\hbar,M)&=\frac{(-1)^\ell}{2\pi^2 k}\Bigg[-4\sum_{m\geq 1,\atop m\neq { \ell k\over 4}}A_m(\hbar,M)\mathcal{I}_1(4m/k-\ell)  +\sum_{m\geq 1, \atop m\neq { \ell k\over 4}}B_m(\hbar,M)\mathcal{I}_0(4m/k-\ell)\\
&+4 \mathcal{I}_2(-\ell)+\frac{1}{2}  \alpha(k,M)\mathcal{I}_0(-\ell) +\delta_{\frac{\ell k}{4},\mathbb{N}}\left(4A_{\frac{\ell k}{4}}(\hbar,M)\left(\frac{E_0^2}{2}-\frac{\pi^2}{6}\right)-B_{\frac{\ell k}{4}}(\hbar,M)E_0\right)\Bigg]~.
\end{split}
\eeq{}
The Dirac delta $\delta_{\frac{\ell k}{4},\mathbb{N}}$ takes the value one when the positive number ${\ell k \over 4}$ takes integer values, otherwise it is zero. We would now like to check whether 
\beq
A(k,M)=\widetilde{A}(k,M)
\eeq{}
and if the coefficients in $J^{{\np}}(\mu,k,M)$ match with $\eqref{jfink}$. In table \ref{Afinktable} we list approximative values of the function $\widetilde{A}(k,M)$, including up to second order exponentially small terms in $\eqref{enexp}$. There are two sources of error for the value of $\widetilde{A}(k,M)$. First, there is an error due to the approximation of the lowest energy level obtained by neglecting exponentially small corrections of order three and higher in $\eqref{enexp}$. Second, keeping a finite number of terms in the Pad\'e approximant when resumming $\mathcal{R}_0(k,M)$ also introduce a numerical error. The error estimate displayed is the largest of these errors, and in table \ref{Afinktable} the largest error is due to neglecting exponentially small corrections in $\eqref{enexp}$. As we can see, the matching between the exact value $A(k,M)$ and $\widetilde{A}(k,M)$ is within the estimated error of approximation.  
\begin{table}[t] 
\centering
\begin{tabular}{c l c c c c}
\hline
$(k,M,q)$  &	$\widehat{A}(k,M)$ &$-\mathcal{R}^{(q)}_0(k,M)$ & $A(k,M)$ & Difference & Error estimate \\
\hline
$(2,1,32)$ & $0.14763411\ldots$ & $0.1380432$ & $0.28567667\ldots$ & $10^{-6}$ & $10^{-6}$ \\
$(3,1,32)$ & $0.15809673\ldots$ & $0.17358596$ & $0.33168359$\ldots &$10^{-6}$  & $ 10^{-6}$\\
\hline
\end{tabular}
\caption{Comparison between the functions $\widetilde{A}(k,M)$ and $A(k,M)$ for $k=2,3$ and $M=1$. The sum of the second and the third column should equal the fourth. In the fifth column the difference is displayed, and in the last column an error estimate. We have included exponentially small corrections up to order two in $\eqref{enexp}$. The value of $q$ is the degree of the Pad\'e approximant used in the Borel resummation. Only stable digits are displayed in the third column.}
 \label{Afinktable}
\end{table}%

Next, using the values of the coefficients $A_\ell(\hbar,M)$ and $B_\ell(\hbar,M)$ given in section \ref{wkbtest} we find 
\beq
\bs
\widetilde{J}^{{\rm np}}(2,1,\mu)=&~\widetilde{C}_1(4\pi,1) \re^{-\mu}+\left[-\frac{4\mu^2 +2\mu}{\pi^2}+\widetilde{C}_2(4\pi,1)\right]\re^{-2\mu}  +\ldots \\
\widetilde{J}^{{\rm np}}(3,1,\mu)=&~\widetilde{C}_1(6\pi,1)\re^{-\mu}-\frac{2}{3}\re^{-4\mu/3}+\widetilde{C}_2(6\pi,1) \re^{-2\mu}-\re^{8\mu/3}+\ldots ~. 
\end{split}
\eeq{jkallfink}
We see that the coefficients that do not depend on $E_0$ match with the ones in $\eqref{jmorifink}$. For the coefficients which do depend on $E_0$, in table \ref{ccoeff} we compare approximate values of these coefficients with the exact values from $\eqref{jmorifink}$. We have included exponentially small corrections up to second order in $\eqref{enexp}$, and the error estimate is due to this truncation. Again we see that the coefficients match within the estimated error.
\begin{table}[t] 
\centering
\begin{tabular}{c l c c c c}
\hline
$(k,M,\ell,q)$ &	$ \widehat{C}_\ell(2\pi k,M)$ &$-\mathcal{R}^{(q)}_\ell(k,M)$ & Exact value & Difference & Error estimate\\
\hline
$(2,1,1,30)$ & $2.45141795\ldots$ & $-2.451437$ & $0$ & $10^{-5}$&$10^{-5}$\\
$(2,1,2,40)$ & $-39.8991984\ldots$ & $39.79831$ & $-\frac{1}{\pi^2}$&$10^{-4}$&$10^{-4}$ \\
$(3,1,1,52)$ & $5.58971619\ldots$ & $-5.589682$ & $0 $ &$10^{-5}$ &$10^{-5}$ \\
$(3,1,2,52)$ & $-157.31222\ldots$ & $157.31093$ & $0$ &$ 10^{-3}$&$ 10^{-3}$ \\
\hline
\end{tabular}
\caption{Comparison between the coefficients in $\eqref{jmorifink}$ and  $\eqref{jkallfink}$. The second and third column should add up to the fourth column. In the fifth column the difference is displayed. We have included exponentially small corrections up to order two in $\eqref{enexp}$. The value of $q$ gives the number of terms kept in the sum when performing the Borel-Pad\'e resummation. Only stable digits are displayed in the third column.}
 \label{ccoeff}
\end{table}%

\section{Summary and outlook}
In this paper we have discussed the spectral problem introduced in \cite{Honda:2014npa} in the context of a Fermi gas formulation of the ABJ matrix model. We have shown strong evidence that the spectral problem can be solved through the WKB quantization condition, with a quantum volume based on expressions from the refined topological string on local $\IP^1\times \IP^1$. These results generalize the ones obtained in \cite{qspec} for the ABJM spectral problem, and we have performed detailed tests of the proposed solution by comparing with numerical values of the energy levels. The solution to the spectral problem allows us to calculate the grand potential of the ABJ model and we have found that the expressions for the grand potential obtained in \cite{Matsumoto:2013nya,Honda:2014npa} are reproduced.

From the formulation of the spectral problem itself, the connection to topological string theory is surprising. It is only knowing that the spectral problem originates from the ABJ matrix model that helps us conjecture the form of the quantum volume. This is different from the spectral problem of the ABJM model. In this case, the spectral problem can be interpreted as a quantization of the mirror curve of local $\IP^1\times\IP^1$, and the connection to topological string theory is clear. Perhaps the same is true also for the ABJ spectral problem, maybe the curve $\eqref{curveabj}$ can be identified with of the mirror curve of local $\IP^1\times\IP^1$ with some clever change of coordinates which we have so far been unable to find.

Spectral problems which are similar to the one studied in this paper appears when studying the M-theory expansion of many different matrix models, see for example \cite{abjmfermi,albanf,Mezei:2013gqa,Kapustin:2010xq}. The spectral problems for these models are much less well understood as compared to the problems studied in \cite{qspec} and in this paper. It would be very interesting to make progress in these cases as well. If we would approach them with the WKB analysis, a first step is to develop techniques for computing the quantum period integrals in a large energy expansion, but for fixed $\hbar$, on the constant energy curve in phase space, corresponding to the curve $\eqref{abjmcurve}$ for the ABJM model. Perhaps the techniques developed for computing the quantum periods for mirror curves in local Calabi--Yaus in \cite{qgeom,Huang:2014nwa} can be adapted. 

Also, a better understanding of how to obtain the non-perturbative part of the WKB quantization condition is needed. Although the conjecture in \cite{qspec} and in this paper is well motivated in view of the connection with the ABJ(M) matrix model, a first principle derivation of the non-perturbative part is at present not known. A better understanding has become even more desirable in view of the results presented in the recent paper \cite{Huang:2014eha}. In there, spectral problems obtained from quantizing various mirror curves of local Calabi--Yaus are studied. As emphasized in \cite{qspec},  the ABJM spectral problem can be understood as quantizing the mirror curve of local $\IP^1\times \IP^1$, for the slice in the complex structure moduli space defined by $\eqref{z1z1Eq}$. In \cite{qspec} it was also speculated that quantizing other mirror curves in a similar way would lead to spectral problems whose spectrum can be found from a WKB quantization condition based on a similar combination of the standard topological string free energy and the free energy of the refined topological string in the Nekrasov--Shatashvili limit, as for the case in the ABJM spectral problem. However, for the cases studied in \cite{Huang:2014eha}, the non-perturbative part in the WKB quantization condition is \textit{not} given purely by the expression based on the topological string free energy, as in this paper and \cite{qspec}. There are more terms, which does not seem to have its origin in amplitudes in topological string theory. A few of them are determined in \cite{Huang:2014eha} by numerical methods. An analytical understanding of how to the derive the non-perturbative WKB quantization condition for the type of spectral problems arising when quantizing mirror curves and studying the M-theory expansion of various Chern--Simons-matter theories would therefore be important to obtain. Possibly this can be done along the lines of \cite{ZinnJustin:1981dx,ZinnJustin:2004ib,ZinnJustin:2004cg,MR729194,MR1483488,MR1704654}.

Of course it would as well be interesting to find completely different techniques to determine the spectrum. In \cite{Fateev:2009jf} a difference equation similar to $\eqref{diffeqabj}$ for $M=0$ is analyzed with methods different from the ones used in this paper. Also, in \cite{MR2165903} a spectral problem similar to the one in \cite{qspec} and in this paper is solved with a different approach, which perhaps can be relevant also in our case.

We believe that there are many things to be understood about how to solve these kind of spectral problems. Given the importance in the context of the gauge/gravity duality we hope to make progress in the future.

\section*{Acknowledgements}
We would like to thank Alba Grassi, Masazumi Honda, Kazumi Okuyama and especially Marcos Mari\~no for inspiring discussions related to the problem studied in this paper. In addition, we are particularly thankful to Yasuyuki Hatsuda for sharing the idea on how to perform the checks in section \ref{coefffink}. We would also like to thank Nordita in Stockholm, Sweden, for hospitality during the course of this work. This work is supported in part by the Fonds National Suisse, subsidies 200020-137523.  
\appendix
\section{Polynomial part of the quantum volume} \label{qvolapp}
We want to compute the quantum volume for $M\geq 0$, up to exponentially small terms in $E$ and $E/k$. We will follow the method outlined in section 5.3 in \cite{abjmfermi}, where the calculation is performed for $M=0$; the case of ABJM. An important ingredient for the arguments in \cite{abjmfermi} was that the functions $T(p)$ in $\eqref{abjT}$ and $U(x,0)$ in $\eqref{abjU}$ grows linearly at infinity, up to exponentially small terms. For $M>0$, $U(x,M)$ has the following behavior
\beq
U(x,M)=\frac{x}{2}+\sum_{n\geq 1}\frac{(-1)^{(M+1)n+1}\re^{-nx}}{n}+\sum_{m=-\frac{M-1}{2}}^{\frac{M-1}{2}}\sum_{s\geq 1}\frac{(-1)^{s+1} 2^s}{s}\left[\sum_{n\geq 1}(-1)^n \re^{-\frac{n x}{k}-\frac{2\pi \ri m n }{k}}\right]^s~,
\eeq{}
and we see that the crucial property of linear growth is preserved also for $M>0$. Therefore, the arguments in \cite{abjmfermi} applies also for $M>0$, and we can use the same formula for the quantum volume as in that paper. Doing this we find
\beq
\bs
{\rm vol}(E;\hbar,M)=&4\Bigg[\int_0^E(2E-2U(x,M))\rd x+\frac{\hbar^2}{24}\int_0^{\infty}U''(x,M)\rd x \\
&\int_{0}^{E}(2E-2T(p))\rd p-\frac{\hbar^2}{48}\int_0^{\infty}T''(p)\rd p-E^2\Bigg]+\mathcal{O}(\re^{-E},\re^{-E/k})~.
\end{split}
\eeq{pavelvol}
The integrals over $p$ are identical to the ones in \cite{abjmfermi}, and they give a contribution
\beq
\int_{0}^{E}(2E-2T(p))\rd p-\frac{\hbar^2}{48}\int_0^{\infty}T''(p)\rd p=\frac{3E^2}{2}-\frac{\pi^2}{6}-\frac{\hbar^2}{96}+\mathcal{O}(\re^{-E},\re^{-E/k})
\eeq{}
For the integrals over $x$, let us consider the cases $M$ even and $M$ odd separately.
\subsection{$M$ even}
When $M$ is even, we have 
\beq
\bs
&\int_0^E(2E-2U(x,M))\rd x+\frac{\hbar^2}{24}\int_0^{\infty}U''(x,M)\rd x =\\
&\int_0^E(2E-2U(x,0))\rd x+\frac{\hbar^2}{24}\int_0^{\infty}U''(x,0)\rd x+I^{(0)}_M+I^{(1)}_M~,
\end{split}
\eeq{qvolcalc1}
where $I^{(0)}_M$ and $I^{(1)}_M$ are given by
\beq
\bs
I^{(0)}_M&=2\sum_{m=-\frac{(M-1)}{2}}^{\frac{M-1}{2}}\int_0^E\log{\left(\tanh{\left(\frac{x}{2k}+\frac{\ri \pi m}{k}\right)}\right)}\rd x \\
I^{(1)}_M&=-\frac{\hbar^2}{24}\sum_{m=-\frac{(M-1)}{2}}^{\frac{M-1}{2}}\int_0^\infty \frac{\rd^2}{\rd x^2}\log{\left(\tanh{\left(\frac{x}{2k}+\frac{\ri \pi m}{k}\right)}\right)} \rd x~.
\end{split}
\eeq{I0I1} 
The first two terms on the RHS in $\eqref{qvolcalc1}$ is the contribution calculated in \cite{abjmfermi}. They are given by
\beq
\int_0^E(2E-2U(x,0))\rd x+\frac{\hbar^2}{24}\int_0^{\infty}U''(x,0)\rd x=\frac{3E^2}{2}-\frac{\pi^2}{6}+\frac{\hbar^2}{48}+\mathcal{O}(\re^{-E},\re^{-E/k})~.
\eeq{}
To calculate the integral in the first line of $\eqref{I0I1}$ we make the change of variables
\beq
t=\re^{-\frac{x}{k}-\frac{2\pi \ri m}{k}}~.
\eeq{}
We then find that
\beq
\bs
&\int_0^E\log{\left(\tanh{\left(\frac{x}{2k}+\frac{\ri \pi m}{k}\right)}\right)}\rd x = - k \int\limits_{\re^{-\frac{2\pi \ri m}{k}}}^{\re^{-\frac{E}{k}-\frac{2\pi \ri m}{k}}}\log{\left(\frac{1-t}{1+t}\right)}\frac{1}{t} \rd t \\
&= -k\left[\Li_2\left(\re^{-\frac{2\pi \ri m}{k}}\right)-\Li_2\left(\re^{2\pi \ri\left(\frac{m}{k}+1/2\right)}\right) +\mathcal{O}(\re^{-\frac{E}{k}})\right]~.
\end{split}
\eeq{}
Next we use the identity
\beq
\Li_2\left(\re^{2\pi \ri x}\right)+\Li_2\left(\re^{-2\pi \ri x}\right)=2\pi^2 B_2(x)~,
\eeq{}
where $B_n(x)$ are the Bernoulli polynomials. For real x, this identity is valid for 
\beq
0\leq x < 1~.
\eeq{}
We therefore have
\beq
\bs
I^{(0)}_M&=-2k \sum_{m=-\frac{(M-1)}{2}}^{\frac{M-1}{2}}\left[\Li_2\left(\re^{-\frac{2\pi \ri m}{k}}\right)-\Li_2\left(\re^{2\pi \ri\left(\frac{m}{k}+1/2\right)}\right) \right] +\mathcal{O}(\re^{-\frac{E}{k}}) \\
&=-4\pi^2 k\sum_{m=\frac{1}{2}}^{\frac{M-1}{2}}\left[B_2\left(\frac{m}{k}\right)-B_2\left(\frac{m}{k}+\frac{1}{2}\right)\right]+\mathcal{O}(\re^{-\frac{E}{k}}) =\frac{\pi^2}{2}(M-k)M+\mathcal{O}(\re^{-\frac{E}{k}}) ~.
\end{split}
\eeq{}
For  $I^{(1)}_M$, we find
\beq
I^{(1)}_M=-\frac{\hbar^2}{24}\sum_{m=-\frac{(M-1)}{2}}^{\frac{M-1}{2}}\frac{1}{k \sinh{\frac{2\ri \pi m}{k}}}+\mathcal{O}(\re^{-\frac{E}{k}})=\mathcal{O}(\re^{-\frac{E}{k}})~.
\eeq{}Adding everything up, we find that the quantum volume for $M\geq 0$ is given by
\beq
\rm{vol}(E;\hbar,M)=8 E^2 -{4 \pi^2 \over 3} +{\hbar^2 \over 24} +2\pi^2\left(M-\frac{\hbar}{2\pi}\right)M +\mathcal{O}(\re^{-E},\re^{-E/k})~.
\eeq{volABJ}
\subsection{$M$ odd}
For $M$ odd, the potential is given by
\beq
U(x,M)=\log{\left(2 \sinh{\frac{x}{2}}\right)}-\sum_{m=-\frac{(M-1)}{2}}^{\frac{M-1}{2}}\log{\left(\tanh{\left(\frac{x}{2k}+\frac{\ri \pi m}{k}\right)}\right)}~.
\eeq{uodd}
For the lowest order term in the $\hbar$ expansion in $\eqref{pavelvol}$ we have
\beq
\bs
&\int_0^E\left(2E-2U(x,M)\right)\rd x\\
&=\frac{3E^2}{2}+\frac{\pi^2}{3}-\frac{\pi^2 k}{2}+ 2\sum_{m=-\frac{M-1}{2}, \atop m\neq 0}^{\frac{M-1}{2}}\int_0^E \log{\left(\tanh{\left(\frac{x}{2k}+\frac{\pi \ri m}{k}\right)}\right)} +\mathcal{O}(\re^{-E},\re^{-\frac{E}{k}}) \\
&=\frac{3E^2}{2}+\frac{\pi^2}{3}-\frac{\pi^2 k}{2}-4\pi^2 k\sum_{m=1}^{\frac{M-1}{2}}\left[B_2\left(\frac{m}{k}\right)-B_2\left(\frac{m}{k}+\frac{1}{2}\right)\right]+\mathcal{O}(\re^{-E},\re^{-\frac{E}{k}}) \\
&=\frac{3E^2}{2}-\frac{\pi^2}{6}+\frac{\pi^2}{2}(M-k)M+\mathcal{O}(\re^{-E},\re^{-\frac{E}{k}})
\end{split}
\eeq{}
where we have reused the results from the previous section for the integral inside the sum. Let us now discuss the term
\beq
\frac{\hbar^2}{24}\int_0^\infty U''(x,M)\rd x~.
\eeq{}
With a similar calculation as for $M$ even we find that the only term in the sum in $\eqref{uodd}$ that gives a contribution which is not exponentially small is the one with $m=0$. We find
\beq
\bs
\frac{\hbar^2}{24}\int_0^\infty U''(x,M)\rd x=&\frac{\hbar^2}{48}\left[\coth{\frac{x}{2}}-\frac{1}{k\sinh{\frac{x}{k}}}\right]^{\infty}_0+\mathcal{O}(\re^{-E},\re^{-\frac{E}{k}})\\
=&\frac{\hbar^2}{48}+\mathcal{O}(\re^{-E},\re^{-\frac{E}{k}})~.
\end{split}
\eeq{}
Adding everything up, we find that also for $M$ odd we have
\beq
\rm{vol}(E;\hbar,M)=8 E^2 -{4 \pi^2 \over 3} +{\hbar^2 \over 24} +2\pi^2\left(M-\frac{\hbar}{2\pi}\right)M+\mathcal{O}(\re^{-E},\re^{-E/k})~.
\eeq{}
\section{Mellin transform} \label{mellin}
In order to compute the grand potential given the quantum volume we need to compute the integrals
\beq
R_{\sigma}^{(j)}(\mu)=\int^{\infty}_{E_0}\frac{E^j \re^{-\sigma E}}{\re^{E-\mu}+1}\rd E~,~j=0,1,2~,\sigma\in\mathbb{Q}~.
\eeq{}
This can be done analytically using the Mellin transform. Given a function $g(u)$ the Mellin transform $\widehat{g}(s)$ is defined by
\beq
\widehat{g}(s)=\int_0^1 g(u)u^{-s-1}\rd u~.
\eeq{}
To compute $R_\sigma^{(j)}(\mu)$ we take the Mellin transform with respect to the variable
\beq
u=\re^{-\mu}
\eeq{}
and use the fact that
\beq
\int_0^1 u^{n-s-1}\left(\log{u}\right)^m \rd u=-\frac{\Gamma(m+1)}{(s-n)^{m+1}}~.
\eeq{mellinint}
From the pole structure of the Mellin transform $\widehat{R}_\sigma^{(j)}(s)$ we can then read off the original function $R_\sigma^{(j)}(\mu)$. The Mellin transform we need is given by
\beq
\int_0^1 \frac{u^{-s-1}}{\re^{E} u +1}\rd u =\re^{sE}I(s) +\sum_{k\geq 1}\frac{(-1)^k \re^{-k E}}{s+k}~.
\eeq{fermimellin}
where
\beq
I(s)=-\pi\csc{(\pi s)}~.
\eeq{}
Note that the function on the RHS in $\eqref{fermimellin}$ does not have poles at negative values of $s$. Using this result we find that the Mellin transform of $R_\sigma^{(j)}(\mu)$ is given by
\beq
\widehat{R}_\sigma^{(j)}(s)=\mathcal{I}_{j}(\sigma-s)I(s)+\ldots
\eeq{}
where $\mathcal{I}_j(n)$ is defined in $\eqref{cali}$ and we have only written out the terms which are relevant for finding the poles. Let us assume that $\sigma$ is not an integer (we can easily obtain the correct expressions when $\sigma$ is an integer from the end results). The above function has poles at $s=\sigma$ and $s=n$, where $n$ is a non-negative integer. Explicitly, we have
\beq
\bs
\widehat{R}_\sigma^{(0)}(s)&=\frac{\pi \csc{(\pi\sigma})}{s-\sigma}-\sum_{n\geq 0}(-1)^n\frac{\mathcal{I}_0(\sigma-n)}{s-n}+\ldots   \\
\widehat{R}_\sigma^{(1)}(s)&=-\frac{\pi \csc{(\pi\sigma})}{(s-\sigma)^2}+ \frac{\pi^2\cot{(\pi \sigma)}\csc{(\pi \sigma)}}{s-\sigma}-\sum_{n\geq 0}(-1)^n\frac{\mathcal{I}_1(\sigma-n)}{s-n}+\ldots \\
\widehat{R}_\sigma^{(2)}(s)&=\frac{2\pi \csc{(\pi\sigma})}{(s-\sigma)^3}- \frac{2\pi^2\cot{(\pi \sigma)}\csc{(\pi \sigma)}}{(s-\sigma)^2}+\frac{\pi^3\csc{(\pi \sigma)}+2\pi^3\cot^2{(\pi\sigma)}\csc{(\pi \sigma)}}{s-\sigma} \\
&-\sum_{n\geq 0}(-1)^n\frac{\mathcal{I}_2(\sigma-n)}{s-n}+\ldots  ~.
\end{split}
\eeq{}
Using $\eqref{mellinint}$ we find that
\beq
\bs
R_\sigma^{(0)}(\mu)&=-\pi \csc{(\pi \sigma)}\re^{-\sigma \mu}+\sum_{n\geq 0}(-1)^n \mathcal{I}_0(\sigma-n)\re^{-\sigma \mu} \\
R_\sigma^{(1)}(\mu)&=-\pi \csc{(\pi \sigma)}\left[\mu+\pi\cot{(\pi\sigma)}\right]\re^{-\sigma \mu} +\sum_{n\geq 0}(-1)^n \mathcal{I}_1(\sigma-n)\re^{-\sigma \mu} \\
R_\sigma^{(2)}(\mu)&=-\pi \csc{(\pi \sigma)}\left[\mu^2+2\pi\mu\cot{(\pi\sigma)}+\pi^2 +2\pi^2\cot^2{(\pi\sigma)}\right]\re^{-\sigma \mu} \\
& +\sum_{n\geq 0}(-1)^n \mathcal{I}_2(\sigma-n)\re^{-\sigma \mu} ~.
\end{split}
\eeq{}
When $\sigma$ takes integer values, the trigonometric functions in the above expressions have poles. These poles are canceled by the poles of the function $\mathcal{I}_j(\sigma-n)$, so the above expressions are valid also for $\sigma$ being an integer, after cancelation of the poles.

\bibliographystyle{utphys}
\providecommand{\href}[2]{#2}\begingroup\raggedright\endgroup

\end{document}